%% file: fs_env_croom_v7.tex
\documentclass[useAMS,usenatbib,letterpaper]{mn2e}
\usepackage{amsmath}
\usepackage{amsfonts}
\usepackage{amssymb}

\usepackage{graphicx,epsfig}
\usepackage{multirow}

\def \oiii {[O{\small~III}]}

\def \ha {H$\alpha$}
\def \hb {H$\beta$}

\def \update {}

\title[GAMA: The environments of HERGs and LERGs]{Galaxy And Mass Assembly (GAMA): The environments of high- and low- excitation radio galaxies}

\input{authors.tex}
\pagerange{\pageref{firstpage}--\pageref{lastpage}}

\begin{document}
\maketitle
\label{firstpage}

\begin{abstract}
We study the environments of low- and high- excitation radio galaxies (LERGs and HERGs respectively) in the redshift range $0.01 < z < 0.4$, using a sample of 399 radio galaxies and environmental measurements from the Galaxy And Mass Assembly (GAMA) survey.  In our analysis we use the fifth nearest neighbour density ($\Sigma_{5}$) and the GAMA galaxy groups catalogue (G3Cv6) and construct control samples of galaxies matched in {\update stellar mass and colour} to the radio-detected sample.

We find that LERGs and HERGs exist in different environments and that this difference is dependent on radio luminosity. High-luminosity LERGs ($L_{\rm NVSS} \gtrsim 10^{24}$ W Hz$^{-1}$) lie in much denser environments than a matched radio-quiet control sample (about three times as dense, as measured by $\Sigma_{5}$), and are more likely to be members of galaxy groups ($82^{+5}_{-7}$ percent of LERGs are in GAMA groups, compared to $58^{+3}_{-3}$ percent of the control sample). In contrast, the environments of the HERGs and lower luminosity LERGs are indistinguishable from that of a matched control sample.   Our results imply that high-luminosity LERGs lie in more massive haloes than non-radio galaxies of similar stellar mass and colour, in agreement with earlier studies \citep{wake08,donoso10}. When we control for the preference of LERGs to be found in groups, both high- and low- luminosity LERGs are found in higher-mass haloes ($\sim 0.2$ dex; at least 97 percent significant) than the non-radio control sample. 
\end{abstract}

\begin{keywords}
surveys -- galaxies: groups: general -- radio continuum: galaxies
\end{keywords}

\section{Introduction}\label{sec:intro}

There is an intriguing interplay between the evolution of a galaxy and
its environment. Evidence that galaxies are influenced by their
surrounding environment includes the observed correlation between star
formation rate (SFR) and local galaxy density for galaxies in clusters
\citep[e.g.][]{lewis02,wijesinghe12,burton13}, along with the
well-known morphology-density relation \citep{dressler80}. 

Feedback between galaxies and their environment can occur in several
ways, but one important process for massive galaxies in the
low-redshift ($z<1$) Universe is {\it radio-mode feedback}
\citep{croton06,bower06}. Mechanical energy is deposited into the
interstellar medium by radio jets and lobes that are powered by an active
nucleus.  The lobes rise and expand due to buoyancy. Unlike supernova-driven feedback, radio-mode feedback can 
provide sufficient energy \citep[$\sim 10^{35}-10^{38}$
W;][]{birzan04} to suppress star formation in massive galaxies without
also requiring a starburst to drive the feedback. 

The jets of radio galaxies can also influence the wider cluster
environment by displacing and heating the gaseous intracluster medium
(ICM) \citep{mcnamara00,mcnamara05}.  This heating of the ICM by radio
galaxies \citep{mcnamara07} can solve the long-standing {\it cooling
  flow problem} \citep{fabian94} where the observed cooling rates
of the ICM are lower than the predicted cooling rates (hundreds to
thousands of solar masses a year) within these systems.  Galaxy
formation simulations can successfully reproduce the observed galaxy
luminosity function if they include this radio-mode heating of the
ICM, whereas an over-abundance of luminous galaxies is produced
without feedback \citep{croton06,bower06}. 

Several properties of radio galaxies are seen to depend on their
environment.  \cite{fanaroff74} type I (FRI) radio galaxies are
dominated by radio emission close to the nucleus, while Fanaroff-Riley
type II (FRII) radio galaxies are dominated by emission from the
lobes.  FRI host galaxies are
found in richer clusters than FRIIs \citep{prestage88,zirbel97}, and
are often the dominant or central galaxy of a cluster.  In contrast, FRII host
galaxies are usually found in smaller groups \citep{zirbel97}.
Related to this, \cite{hine79} found that radio galaxies in clusters
(typically FRIs) tended to show only absorption lines or weak
[O{~\small II}] emission in their optical spectra, whereas most radio
galaxies with strong emission lines were not members of clusters.
This is not surprising given the known star formation--density
\citep{lewis02} and morphology--density \citep{dressler80} relations. 

In our current picture \cite[e.g.][]{hardcastle07,best12}, the
observed optical spectrum of a radio galaxy reflects the presence (or
absence) of a radiatively efficient accretion disk surrounding the
central supermassive back hole (SMBH).  To form such a disk the AGN
requires a sufficiently high accretion efficiency (typically
1--10 percent of the
Eddington rate).  The UV radiation field from the disk can
then ionize surrounding gas leading to the presence of high-excitation
emission lines.  Objects with such features are termed high-excitation
radio galaxies, or HERGs.  At lower accretion efficiency (less than
$\sim1$ percent of Eddington) a radiatively inefficient advection dominated accretion flow
(ADAF) can form.  The lack of UV
continuum in this case leads to a dearth of strong emission lines.
Such objects are termed low-excitation radio galaxies, or LERGs.

While the observational characteristics of HERGs and LERGs are
primarily driven by the accretion rate onto the black hole, the flow
of gas from larger scales is expected to modulate this.  Cold gas can
be quickly accreted from galaxy scales to the scale of the black hole,
particularly in the presence of dynamical disturbance generated by
interactions or mergers.  This ``{\it cold-mode accretion}'' can be
rapid enough to enable the formation of a radiatively 
efficient accretion disk, which will then photoionize the surrounding
gas and produce strong, high-excitation optical emission lines
(i.e. HERGs).  In contrast, high temperature gas (e.g. gas
shock-heated to the virial temperature of the host halo of the galaxy)
in massive galaxies will cool only slowly, providing a slow trickle of
gas onto the black hole, insufficient for a radiatively efficient
disk.  As a result such objects would have weak or non-existent
emission lines (LERGs).

These two distinct populations of high-excitation and low-excitation
radio galaxies may provide feedback in different ways.  The LERGs are
postulated to lie in high-mass quasi-static hot halos, which supply
cooling gas onto the radio galaxy, and the accretion process is
self-regulated by mechanical feedback from the radio jets, i.e. ``{\it
  radio-mode}'' feedback
\citep[e.g.][]{keres05,croton06,bower06,van-de-voort12}.  In contrast,
HERGs are thought to have radiative and mechanical feedback from both
the AGN and supernovae in the host galaxy, which is termed ``{\it
  quasar-mode}'' feedback \citep{croton06,bower06}.  

Early studies of radio galaxy environments were focused on the most
massive bound systems, i.e. clusters or massive groups.  In part this
was due to the difficulty of finding lower mass groups in relatively
shallow galaxy samples.  New surveys, such as the  Sloan Digital Sky Survey
\citep[SDSS;][]{york00}, 2-degree Field Galaxy Redshift Survey
\cite[2dFGRS;][]{colless01}, 6-degree Field Galaxy Survey
\citep[6dFGS;][]{jones04} have dramatically increased our ability
study galaxy environments.  The
availability of optical spectra for fainter galaxies has enabled more
detailed studies over a wider range in group masses, by 
enabling individual group members to be robustly identified. 

In this paper we use the Large Area Radio Galaxy Evolution
Spectroscopic Survey \citep[LARGESS;][]{2017MNRAS.464.1306C} to carry out new
investigations into the impact of environment on radio galaxies.
LARGESS contains a total of 10,856 spectroscopically confirmed radio
galaxies.  The LARGESS radio
galaxies are combined with data from the Galaxy And Mass Assembly survey
\citep[GAMA;][]{driver11} and we specifically focus on the differences
between the environments of high-- and low--excitation
radio galaxies, making use of the highly complete GAMA group catalogue
\cite{robotham11}.  We describe the GAMA galaxy and radio galaxy samples
in \S\ref{sec:env_sample}, and explain our classification scheme for 
low-- and high--excitation radio galaxies and stellar mass estimates in
\S\ref{sec:env_galproperties}. \S\ref{sec:stat_ana} summarises the statistical methods used. In \S\ref{sec:d5} we present results based on the fifth
nearest neighbour density of radio galaxies.  In \S\ref{sec:fof} 
we present the group occupation statistics for the sample using the GAMA groups catalogue and then investigate detailed group properties in \S\ref{sec:fof_prop}.
We provide a detailed discussion of our
results in the context of previous work in \S\ref{sec:env_discussion}
and summarise our results in \S\ref{sec:env_summary}. 

Unless otherwise stated, we assume: $H_{0}$ = 100 $h$ km s$^{-1}$
Mpc$^{-1}$, $\Omega_{\Lambda}=0.75$ and $\Omega_{m}=0.25$, which
corresponds to the cosmology of the Millennium N-body simulation used
to construct the GAMA light cone mocks and GAMA groups catalogue. The
original cosmology used in all other GAMA measurements was $h =
0.7$, $\Omega_{\Lambda}=0.7$ and $\Omega_{m}=0.3$.  In our results we
will factor out the $h$ dependence and assume the difference in
$\Omega_{\Lambda}$ and $\Omega_{m}$ are insignificant at the
relatively low redshifts of our sample. 

\begin{table*}
\begin{center}
\caption{The number of LARGESS targets, observed objects and objects with good redshifts ($nQ>2$) that lie within the GAMA survey area. The number of filler targets that are included only because of a FIRST detection is shown in parenthesis. The number of objects with a reliable redshift (internal data: SpecObjv08) is based on the GAMA redshift flag ($\emph{nQ}>2$). ``Radio limit'' refers to the number of objects that have a NVSS match and FIRST flux density greater than 3.5 mJy.}
\label{tab:obs}
\begin{tabular}{l rr rr rr}
\hline
 & \multicolumn{2}{c}{Targets} & \multicolumn{2}{c}{Observed} & \multicolumn{2}{l}{Redshift ($\emph{nQ}>2$)}\\
Sample              &                       All&Radio limit & All&Radio limit & All & Radio limit\\
\hline
Main GAMA target & 1,470&599 &  1,470&599 &  1,433&582\\
Filler target (all) & 1,624&756 &  1,319&596 &   993&458 \\ 
Filler target (radio) & (1,442)&(691) &  (1,155)&(539) &   (850)&(409) \\ 
\hline
Total                 &  3,094&1,355 &  2,789&1,195 &  2,426&1,040\\
\hline 
\end{tabular}
\end{center}
\end{table*}

\section{Sample and observations}\label{sec:env_sample}

\subsection{The GAMA Survey}

The GAMA survey \citep{driver11,hopkins13} is a large multi-wavelength
study of galaxy formation and evolution using the multi-object
2-degree Field (2dF) robotic fibre positioner, coupled with the
AAOmega spectrograph \citep{2006SPIE.6269E..14S} on the Anglo-Australian Telescope (AAT). 2dF has
392 fibres (and 8 guide fibres), each with a diameter of
2\,arcsec on the sky (corresponding to $\sim9$ kpc at $z$=0.3). The
main GAMA I spectroscopic survey targeted galaxies that were photometrically selected from the SDSS sixth data
release \citep{adelman-mccarthy08} in three $4\times12$ degree
equatorial regions at approximately 09h, 12h and 15h in right
ascension.  The GAMA II sample \citep{2015MNRAS.452.2087L} extends the area of the GAMA I regions
and increases the depth of the spectroscopy across the whole survey to
a limiting magnitude of $r_{pet} < 19.8$ (extinction corrected).
Galaxies which already had spectra from SDSS (or other sources) were
in general not re-observed as part of GAMA.  In this work we will only consider galaxies within the GAMA I sky coverage but we will use data products from both GAMA I and II, i.e. those to a limiting depth of $r_{\rm pet}=19.8$. 

For each GAMA spectrum, a heliocentric redshift was measured using the program \textsc{runz} \citep{saunders04,drinkwater10}.  \textsc{runz}  cross--correlates with template spectra, fits emission lines and allows manual checking and interaction.  As part of the manual checking of the redshift, a quality flag \cite[{\em Q};][]{driver11} was assigned.  Redshifts for radio galaxy targets were also independently re-derived with a template set that included broad--line AGN (that were not included in the main GAMA redshift pipeline).  The broad line AGN template used was an SDSS quasar composite spectrum \citep{2001AJ....122..549V}.  Approximately 8 percent of redshifts and classifications disagreed between LARGESS and GAMA, with most cases being broad--line AGN at high redshift.  As part of the quality control
process for GAMA redshifts, several GAMA team members independently checked the redshift and quality values \citep{2015MNRAS.452.2087L}.  Final values were then assigned based on the combined redshifts and qualities ({\em nQ}).  In this paper we only use radio galaxies where the redshift difference is within 0.01 (i.e. $|z_{\rm LARGESS} - z_{\rm GAMA}| < 0.01$) and both surveys have considered the redshift to be reliable ($\emph{nQ}> 2$).

\begin{figure}
\centering
\includegraphics[width=0.50\textwidth, trim=20 0 0 0]{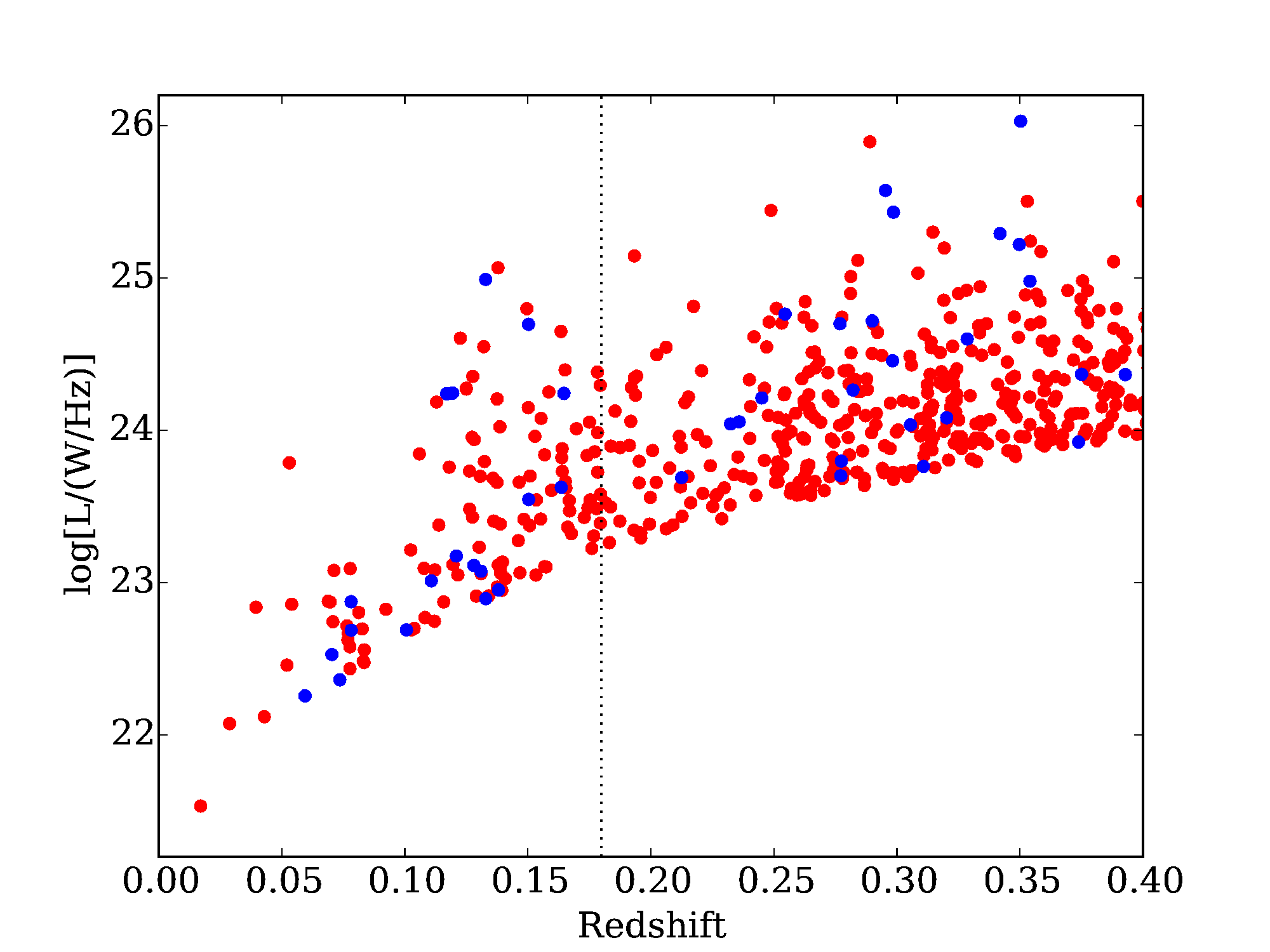}
\caption{The distribution of radio luminosity and redshift for our sample.  LERGs and HERGs are indicated by red and blue points respectively.  The dotted line at $z=0.18$ marks the upper redshift limit for the 5th nearest neighbour analysis.}
\label{fig:lz}
\end{figure}

\subsection{The radio galaxy sample}

A full description of the LARGESS sample is given by \citet{2017MNRAS.464.1306C}.  For completeness, we provide a brief outline here.  The LARGESS sample comprises 19,179 radio-optical matches covering over 800 deg$^2$. Radio sources are identified by a cross-match between the SDSS down to an extinction corrected $i_{\rm model} < 20.5$ and the Faint Images of the Radio Sky at Twenty-cm \cite[FIRST;][]{becker95} to 5 times the local root mean-square noise ($\simeq1$ mJy). The cross-matching follows a tiered-algorithm, and is optimised to improve the identification of extended radio sources at the chosen depth.  The sample was also matched to the NRAO VLA Sky Survey \cite[NVSS;][]{condon98} to get good estimates of total flux for extended sources.  The overall optical-radio matching reliability and completeness of the sample are estimated to be 93.5 percent and 95 percent respectively \citep[see][]{2017MNRAS.464.1306C}.  For objects that did not already have high quality spectra from SDSS we obtained followup spectroscopy as part of the GAMA and WiggleZ \citep{drinkwater10} surveys.  LARGESS sources that were not part of the main GAMA or WiggleZ samples were targeted at lower priority.  The final catalogue contains spectroscopic observations for 12,329 radio sources, of which 10,856 have reliable redshifts.

In this paper we limit our radio galaxy sample to those LARGESS objects in the GAMA I regions. Out of the full 19,179 LARGESS target sample, 3,168 lie within the GAMA I spectroscopic survey coverage. Of these, 3,094 fall within the area covered by the GAMA GAMA I spectroscopic tiling catalogue (internal data management unit: InputCatv16). 1,470 of these were in the main GAMA I sample \citep[internal flag: {\tt survey\_class} $\geqslant 4$, see][]{driver11} and the remaining 1,624 were filler targets, which are mostly targets only because of a FIRST detection [$\simeq180$ were targeted as fillers for other reasons \citep[see][]{baldry10}].  Spectra were obtained for $\sim80\%$ of our `filler' targets (fainter than $r_{\rm pet}=19.8$), more than 70\% of which have a reliable redshift (see Table \ref{tab:obs}).  Finally, we also limit our sample to LARGESS galaxies with a NVSS detection and FIRST total integrated flux density ($S^{\rm FIRST}_{tot}$) above 3.5 mJy. The NVSS measurements provides better surface brightness sensitivity to extended emission. The FIRST limit is applied because the fraction of LARGESS objects with a NVSS match drops rapidly below this limit.  The distribution of objects in the radio luminosity vs.\ redshift plane is shown in Fig.\ \ref{fig:lz}.  Radio luminosities, here and elsewhere, are calculated using {\update our default cosmology (with $h=1$)}, and a K-correction assuming a power law spectral index of $\alpha=-0.7$.  As expected, given a flux limited sample, the typical luminosity varies across the volume we sample.    To counteract this we make comparisons to control samples throughout our analysis.  The 5th nearest neighbour analysis is carried out within a volume limited to $z<0.18$ and we use control galaxies with the same redshift distribution for our group analysis.

\section{Measured galaxy properties}\label{sec:env_galproperties}

\subsection{Low-- and high--excitation galaxy classification}\label{sec:env_class}

\begin{figure}
\centering
\includegraphics[width=0.50\textwidth, trim=50 0 0 0]{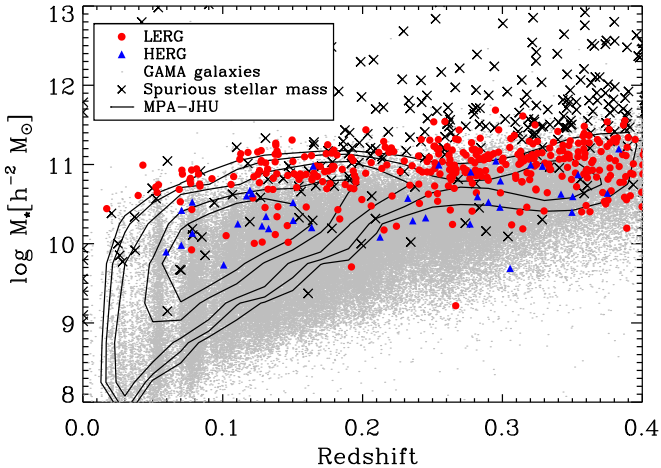}
\caption[The stellar mass as a function of redshift for a subset of the full GAMA II sample and radio galaxies]{The stellar mass as a function of redshift for a subset ($z<0.4$) of the full GAMA II sample (grey points) and radio galaxies coloured according to their final classification: LERG (red) and HERG (blue). Contours of SDSS galaxies with stellar mass estimates from MPA-JHU are shown for comparison. The crosses mark GAMA galaxies with spurious stellar masses (see \S\ref{subsec:mstar_est}), which we remove from our analysis.}
\label{fig:massVz}
\end{figure}

We classified galaxies into four main categories, star-forming (SF) galaxies, low-excitation radio galaxies (LERGs), high-excitation radio galaxies (HERGs) and broad emission line galaxies (AeB). The full details are provided by \citet{2017MNRAS.464.1306C} but we give a brief summary below.

Broad emission line objects were classified by visual inspection of GAMA, WiggleZ and some SDSS spectra, and by spectroscopic flags from the surveys (particularly SDSS). Galaxies classified as AeB are not included in our analysis because the non-stellar continuum contribution from the accretion disk biases galaxy parameters such as stellar mass.  In the unified model of high--accretion--rate AGN \citep{urry95}, we expect AeB objects to have the same intrinsic properties as HERGs.  A difference in orientation leads to the observational differences.

The radio luminosity function above $10^{24}$\,W\,Hz$^{-1}$ is dominated by AGN \citep[e.g.][]{mauch07,best12}, with a space density of $\sim4\times10^{-6}$ Mpc$^{-3}$ for luminosities between $10^{24.1}$--$10^{24.7}$\,W\,Hz$^{-1}$, compared to just $\sim 6\times10^{-8}$ Mpc$^{-3}$ for star-forming galaxies over the same luminosity range. The decline of SF galaxies above $10^{24}$ W Hz$^{-1}$ is because the inferred SFR exceeds $\sim 550 M_{\sun}$ yr$^{-1}$ \citep{hopkins03}, and such sources are rare in the low-redshift Universe. Therefore, we considered all galaxies with FIRST radio luminosities ($L_{\rm FIRST}$) greater than $10^{24}$ W Hz$^{-1}$ to have radio emission produced by an AGN.  The H$\alpha$ line is redshifted off the end of the observed spectral range at $z\sim0.35$ for GAMA and at $z\sim0.4$ for the SDSS and WiggleZ.  Our adopted FIRST flux limit of 3.5 mJy is equivalent to a radio luminosity of $L_{\rm FIRST}=10^{24}$ W Hz$^{-1}$ at $z\simeq0.3$, so any object in our sample will be above this luminosity at higher redshift.  We therefore consider all objects at $z>0.3$ to have their radio emission dominated by an AGN.

For objects with $L_{\rm FIRST} \leqslant 10^{24}$ W Hz$^{-1}$ and $z\leqslant0.3$ we used two methods to identify galaxies where the radio emission is likely coming from star formation. The first method used the \citet*[][hereafter BPT]{baldwin81} emission line diagnostic. In this diagnostic, star-forming galaxies were those with H$\alpha$, H$\beta$, [NII]$\lambda6583$ and [OIII]$\lambda5007$ detected at 3$\sigma$ significance, and in the ``pure'' star forming region defined by \cite{kauffmann03a}. The second method compared the SFR inferred from the FIRST radio flux with the SFR inferred from H$\alpha$ emission line flux \citep{hopkins03} for galaxies with H$\alpha$ and H$\beta$ detected at $\geqslant3\sigma$. Galaxies within the typical $3\sigma$ dispersion of the two measurements were classified as having radio emission associated with star forming regions. All remaining galaxies were considered to have radio emission primarily from an AGN \citep[see][for details]{2017MNRAS.464.1306C}.

We lastly classified each of the AGN as either a HERG or a LERG.  This was done on the basis of their \oiii\ emission.  If an AGN had \oiii\ emission detected at $\geqslant3\sigma$ significance and an \oiii\ equivalent width of $>5$\AA\ it was  classified as a HERG, otherwise it was classified as a LERG.  For a more detailed discussion of the efficacy of approaches to HERG and LERG classification see Pracy et al.\ (2016).

\begin{figure}
\centering
\includegraphics[width=0.5\textwidth, trim=30 0 0 0]{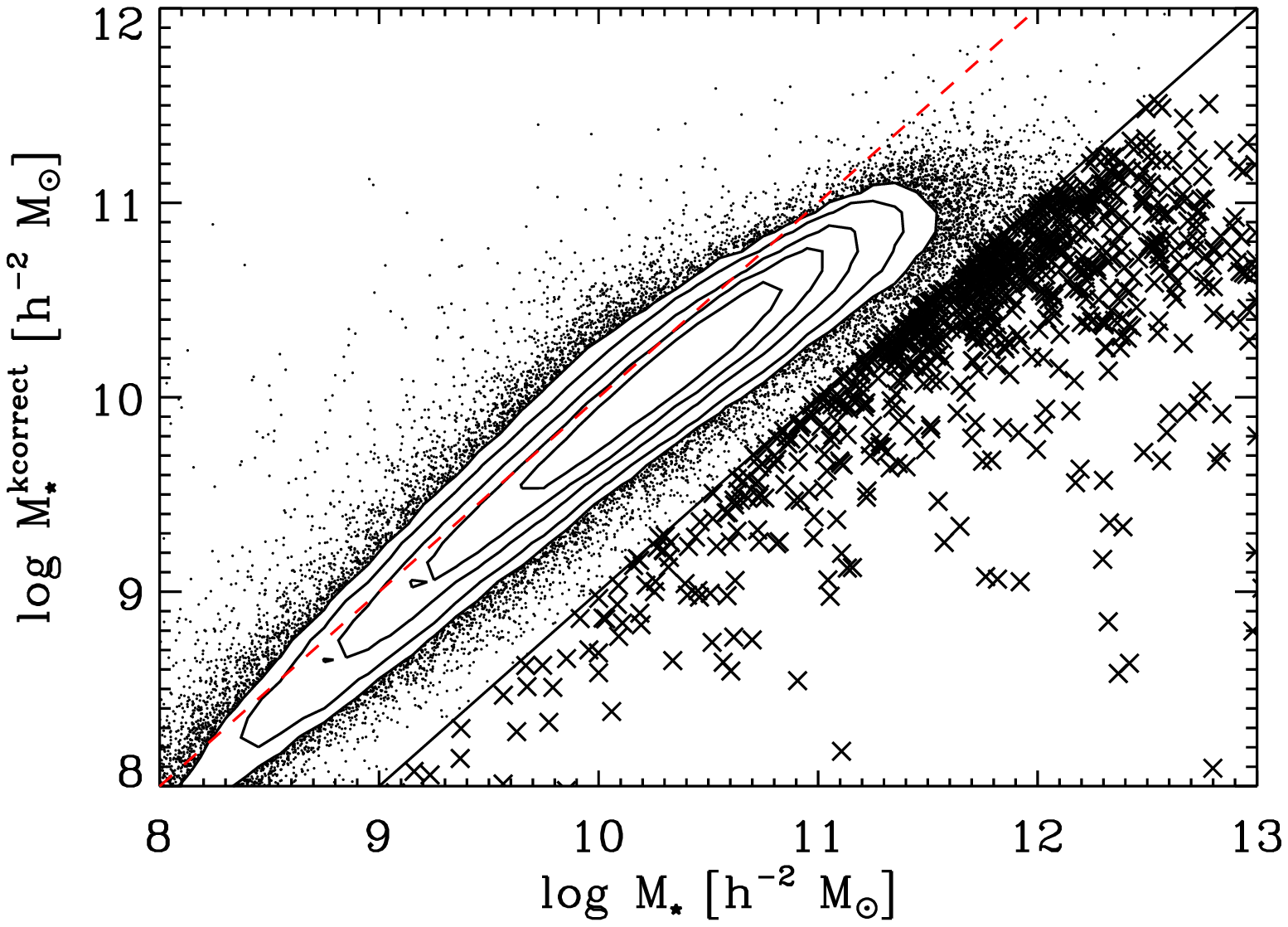}
\caption[Comparison between the stellar masses derived by using the empirical relation in \citet{taylor11} and those derived using the \textsc{kcorrect v4\_2} code]{Comparison between the stellar masses derived by using the empirical relation in \citet{taylor11} (and used in this work) and those derived using the \textsc{kcorrect v4\_2} code \citep{blanton07}. The solid line indicates where ($\log M_{\star} - \log M^{\textsc{kcorrect}}_{\star}=$ 1.0 dex). Galaxies with spurious stellar masses  ($\log M_{\star} -\log M^{\textsc{kcorrect}}_{\star}>$ 1.0 dex; crosses) are removed from our analysis. The one-to-one relation is indicated by the red dashed line.}
\label{fig:massVmass}
\end{figure}

\subsection{Stellar mass estimates}\label{subsec:mstar_est}

\cite{taylor11} made stellar mass estimates using spectral energy distribution (SED) fits to the GAMA photometry of SDSS images. Using these estimates \cite{taylor11} then presented a tight ($\simeq0.1$ dex scatter) empirical relation (equation 8 in their paper) to derive stellar masses using the rest-frame $g-i$ colour, and $i$-band luminosity. We use this relationship (applied to SDSS Petrosian magnitudes) to generate stellar mass estimates for the galaxies in our work. We make our own estimates because the stellar masses estimated using the SED fits are not available for all radio targets.

Radio luminosity is known to depend on the stellar ($M_{\star}$) and/or black hole ($M_{\rm
  BH}$) mass of a galaxy \citep[e.g.][]{fabbiano89, taylor96,
  mclure99, mclure04, best05a, mauch07, janssen12}. Fig.\
\ref{fig:massVz} shows the stellar mass distribution of the full GAMA
sample (grey points) and the radio galaxies (coloured) as a function of
redshift. We also include contours of SDSS galaxies from the
MPA-JHU\footnote{\tt http://www.mpa-garching.mpg.de/SDSS/DR7/} stellar mass estimates \citep{kauffmann03}. Since radio-loud AGNs are generally hosted by very massive galaxies, it is important to account for any bias caused by this dependence on stellar mass.
Fig.\ \ref{fig:massVz} shows that our radio galaxy sample is dominated by galaxies with higher masses than the average GAMA galaxy, particularly for the LERGs (red points).  There are a small number of outliers towards low mass.  The lowest mass LERG outliers are found to have extended radio lobes (and therefore not dominated by star formation in the radio), with significant \ha\ emission, but weak or undetectable \hb\ and \oiii\ emission which leads to a LERG classification.  Inclusion of these objects does not influence our results.

\begin{figure*}
\centering
\includegraphics[width=90mm,trim=20 0 20 0]{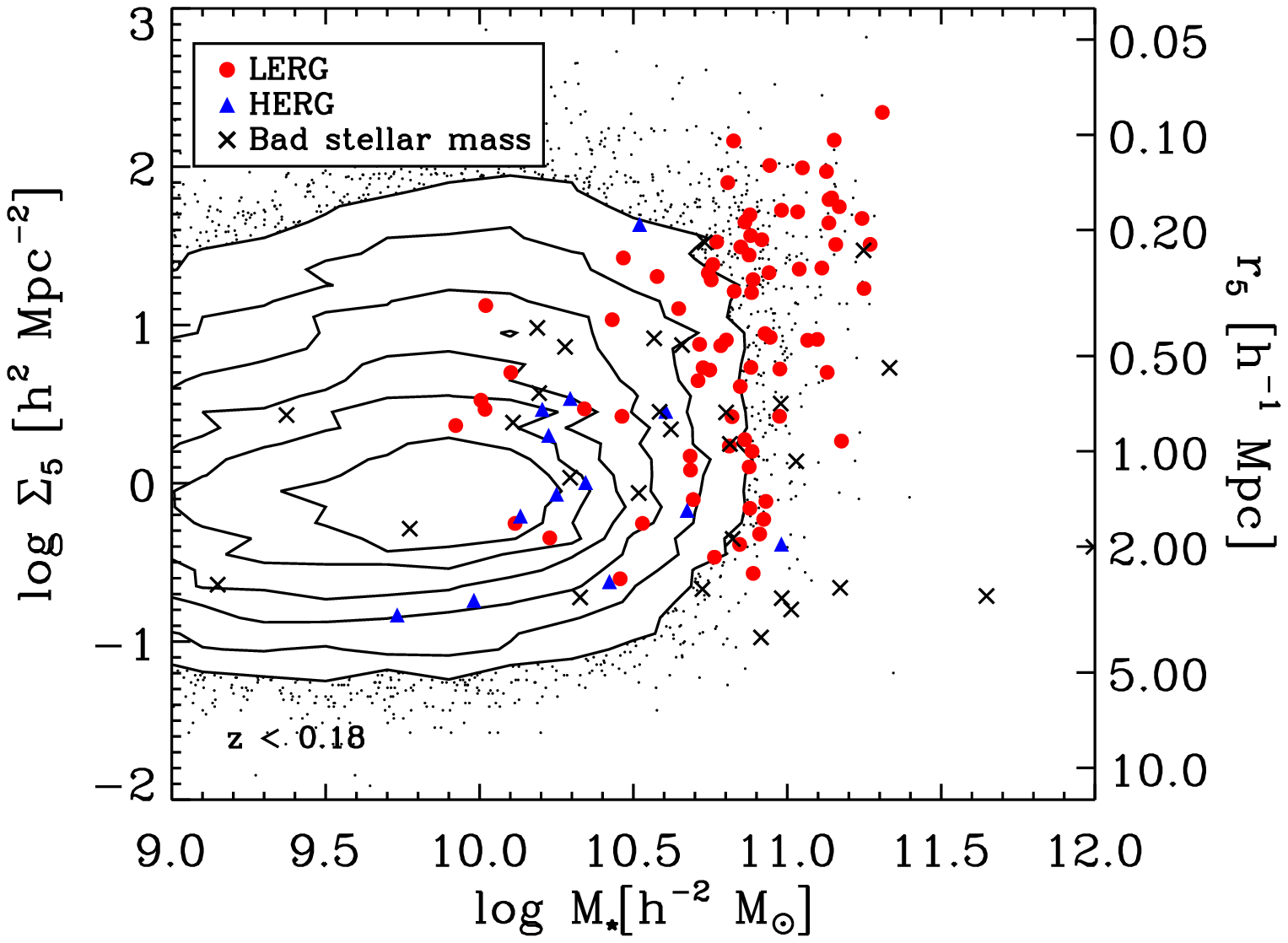}\includegraphics[width=90mm,trim=20 0 20 0]{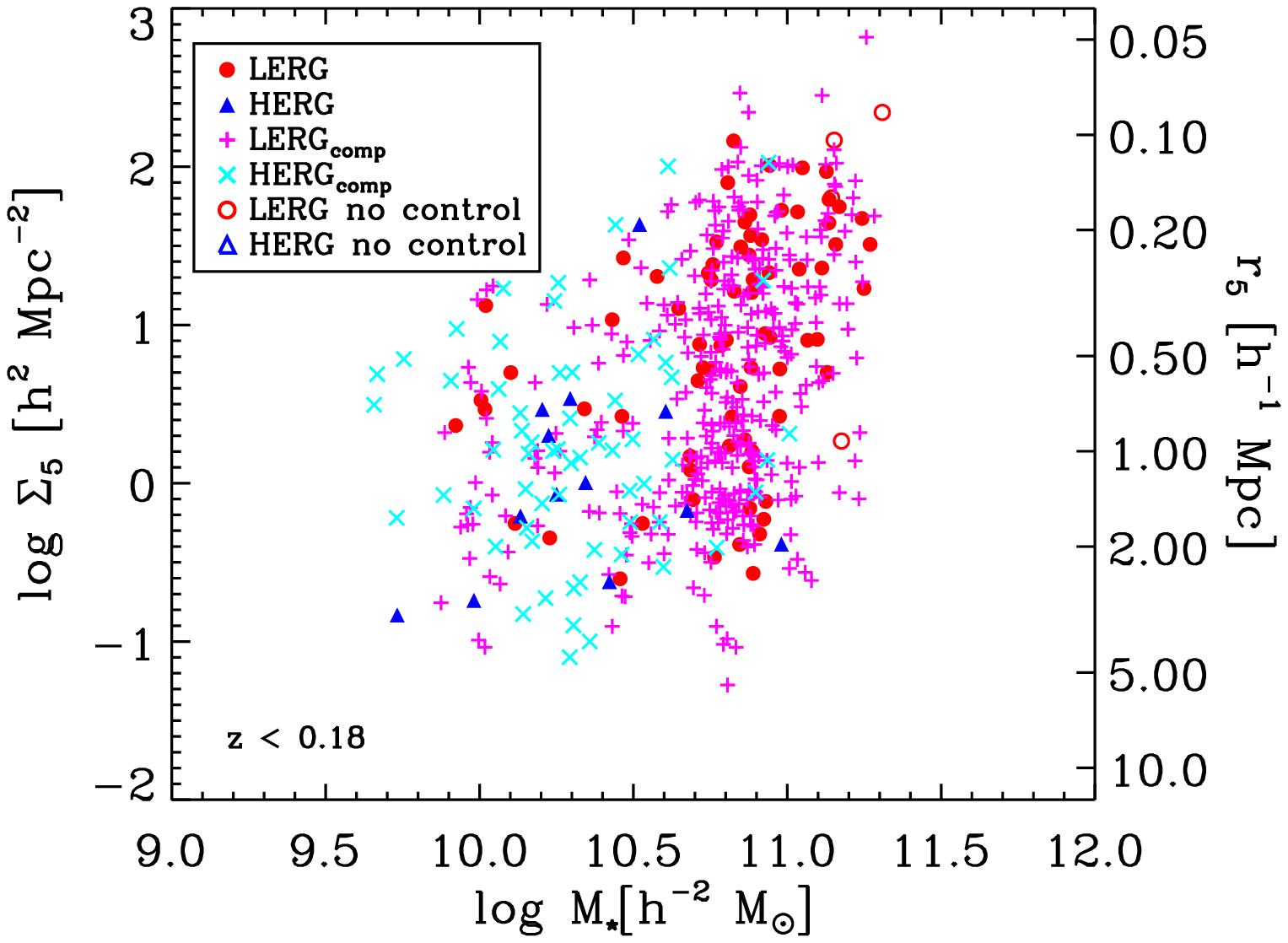}
\caption{The local galaxy density ($\Sigma_{5}$) as a function of stellar mass. Left panel: the GAMA sample in the full redshift range of the density measurements (black points and contours), and radio-loud AGN of different classes (LERG; red filled circle and HERG; blue filled triangle). Also shown are non-radio GAMA galaxies with spurious stellar masses (crosses). Right panel: LERG and HERG samples matched to non-radio control galaxies (LERG$_{\rm comp}$ as magenta plus sign and HERG$_{\rm comp}$ as cyan cross). Open symbols indicate radio galaxies that have been removed because they do not have five non-radio control galaxies.}
\label{fig:denVmass}
\end{figure*}

For GAMA main survey galaxies we compare the \cite{taylor11} stellar masses to those independently derived using
the \textsc{kcorrect v4\_2} code \citep{blanton07} to fit the SDSS $ugriz$ model magnitudes.  Fig.\ \ref{fig:massVmass} shows a tight correlation between the two approaches, but also an offset from the one-to-one line. The stellar masses used in this work are larger than those estimated using \textsc{kcorrect}, which is caused by the different types of photometry \cite[][applied SED fits to GAMA photometry derived using SDSS images and not SDSS derived photometry]{taylor11} and different methods \cite[][used a Bayesian method to estimate the parameters of the stellar population]{taylor11}.  We are not especially concerned by this offset, as the primary use of the stellar masses is to generate matched samples, so as long as we use the same estimate for both the radio galaxies and controls, we should not introduce any bias.  However, there are galaxies that scatter far from the one-to-one line (crosses in Fig. \ref{fig:massVmass}).  Visual inspection of galaxies with spurious stellar masses ($\log M_{\star} - \log M^\textsc{kcorrect}_{\star}>$ 1.0 dex) usually reveals objects or imaging artefacts that affect the photometry.  Because the galaxies with spurious stellar masses (which are $\lesssim$1\% of the total GAMA sample) are scattered towards higher stellar masses (see Figs.\ \ref{fig:massVz} and \ref{fig:massVmass}) this can cause large problems for an analysis of the most massive galaxies, which are uncommon and contain a significant fraction of the radio galaxy population. Therefore we remove all galaxies with spurious stellar masses ($\log M_{\star} - \log M^\textsc{kcorrect}_{\star}>$ 1.0 dex) from our analysis.

\section{Statistical analysis}\label{sec:stat_ana}

We compare the environments of LERGs and HERGs to non-radio control samples drawn from the main GAMA sample and matched to the host galaxy properties of LERGs and HERGs. We also investigate the relationship between the environmental properties and radio luminosity by assigning the radio luminosity of the radio galaxy to its matched non-radio control subset.

There are two basic types of environmental metrics in this analysis, those with a continuous distribution (e.g. fifth nearest neighbour density, halo mass etc.) and those with discrete values (e.g. numbers of galaxies in a group or not). The two cases require their own statistical test to quantify the significance of any observed differences.

\subsection{Continuous quantities}

For environmental metrics with a continuous distribution, we apply a two-sample Kolmogorov-Smirnov test (KS-test) to: (i) the full sample, (ii) galaxies with $L_{\rm 1.4GHz} > L_{\rm 1.4GHz,limit}$, and (iii) galaxies with $L_{\rm 1.4GHz} \leqslant L_{\rm 1.4GHz,limit}$. The $L_{\rm 1.4GHz,limit}$ value is taken to be approximately the median $L_{\rm 1.4GHz}$ of the radio galaxy subset, so there is approximately equal numbers of radio galaxies in each radio luminosity bin. We also calculate the (partial and Spearman rank) correlation coefficient between the environmental metric and the radio luminosity for radio galaxies and their control sample, and compare the two correlation coefficients. The comparison of the correlation coefficients is made by comparing the Fisher transform of the two (radio galaxies and their control sample) correlation coefficients, expressed as a $Z$-test statistic (i.e. $\sigma$ significance) \citep{fisher25}.

\subsection{Discrete quantities}

For environmental metrics that concern discrete quantities (i.e. success rates), we test the significance between the two rates for: (i) the full sample, (ii) galaxies with $L_{\rm 1.4GHz} > L_{\rm 1.4GHz,limit}$, and (iii) galaxies with $L_{\rm 1.4GHz} \leqslant L_{\rm 1.4GHz,limit}$.  We use a Bayesian approach to estimate the significance that two rates are different. \cite{lee06} presented the Bayes factor as a measure of significance for comparing success rates. The Bayes factor for success rates is the ratio between the marginal probability for a model assuming the two rates are the same, $P(D|M_{s})$, and the marginal probability for a model assuming the two rates are different, $P(D|M_{d})$. $P(D|M_{s})$ and $P(D|M_{d})$ are marginal probabilities and so are not normalised, but given that the two models are the only plausible models, we obtain normalised probabilities by marginalising the models i.e. $P_{N}(D|M_{d}) + P_{N}(D|M_{s}) =1$. In this paper we will show the normalised probability of the ``different rate'' model (i.e. $P_{N}(D|M_{d}) = (1+B)^{-1}$, where $B$ is the Bayes factor calculated using the relation in \cite{lee06}) and prefer the different rate model if the probability is greater than or equal to 95\%.

\begin{figure*}
\centering
\includegraphics[width=90mm,trim=20 0 20 0]{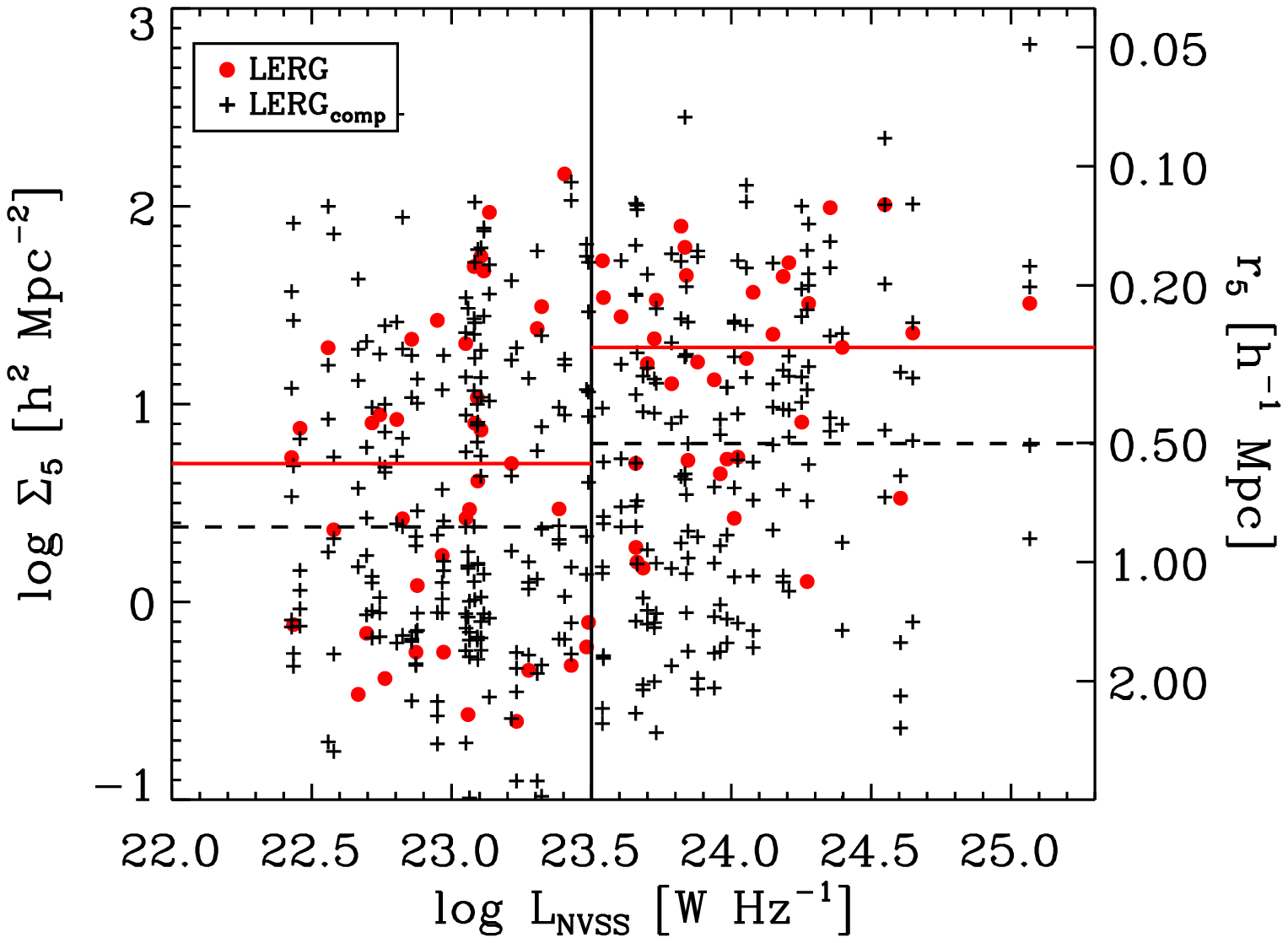}\includegraphics[width=90mm, trim=20 0 20 0]{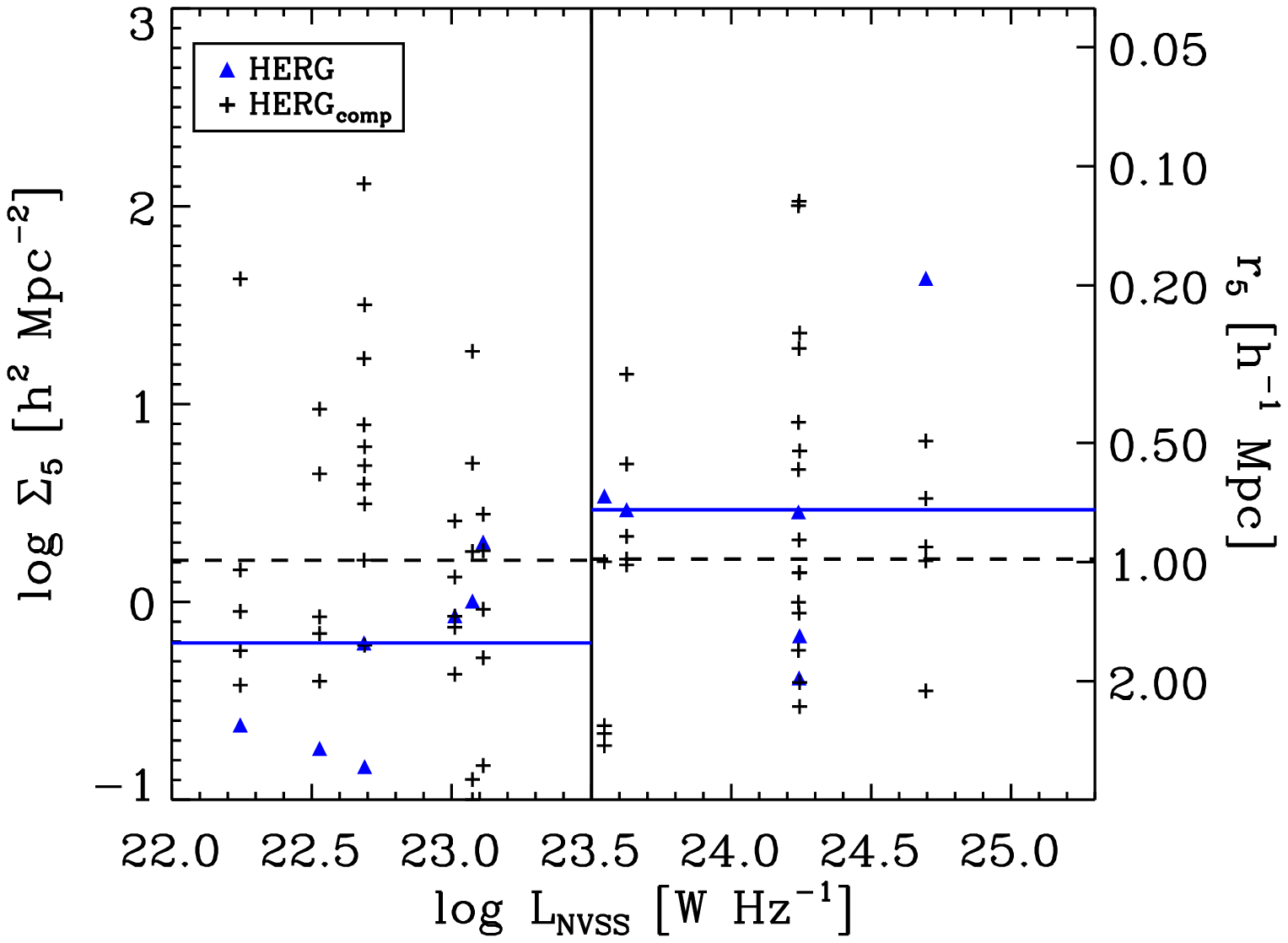}
\caption{The $\Sigma_{5}$ distribution for radio galaxies (LERGS: left panel, red circles and HERGs: right panel, blue triangles) and their control galaxies (plus signs) as a function of radio luminosity. Control galaxies are assigned the radio luminosity of the radio galaxy that they have been matched to.  The horizontal lines are the median $\Sigma_{5}$ for radio galaxies (solid) and their control sample (dashed) in two different radio luminosity bins -- $L_{\rm NVSS} \leqslant 10^{23.5}$ W Hz$^{-1}$ and $L_{\rm NVSS} > 10^{23.5}$ W Hz$^{-1}$.  The displayed control galaxies (and their median values) are chosen from a single random realisation.}
\label{fig:den_dn_match}
\end{figure*}

\section{The fifth nearest neighbour density of radio galaxies}\label{sec:d5}

The fifth nearest neighbour density, $\Sigma_{5}$, uses the projected distance, $r_{5}$, to a galaxy's 5th nearest neighbour, and then converts this to a surface density such that
\begin{equation}
\Sigma_5 = \frac{5}{\pi r^{2}_{5}}.
\end{equation}
{\update $\Sigma_{5}$ is a valuable environmental metric as it is related to dark matter halo density \citep{sabater13}, that is well correlated with halo mass \citep{haas12}.}
We calculated $\Sigma_{5}$ for all the radio galaxies that have redshifts $0.002<z<0.18$.   A maximum radial velocity separation of $\pm1000$ km s$^{-1}$ for the fifth nearest neighbour was applied to limit projection effects \citep[e.g.][]{rowlands12,brough13}. The density defining population used to find the 5-th nearest neighbour was a volume limited subset of GAMA galaxies with $z<0.18333$ and $M_{r} < -20  -Qz$.  The extended range in redshift accounted for the upper velocity range and $Q=0.87$ \citep{loveday12} accounted for the evolution of $M_{r}$ as a function of redshift.  Additionally, we corrected the calculated density for redshift incompleteness in the GAMA sample and we did not use galaxies for which $r_5$ was greater than the projected distance to the nearest edge of the survey area.  Only radio galaxies that are also in the main GAMA sample were included in this analysis.

The left panel of Fig.\ \ref{fig:denVmass} shows $\Sigma_{5}$ against stellar mass for GAMA galaxies with $z<0.18$. As in Fig.\ \ref{fig:massVz}, we can clearly see that LERGs (red points) and HERGs (blue points) cover different ranges in stellar mass, but this figure also shows the relative excess of LERGs in high-density environments. The LERGs and HERGs span a relatively narrow range in stellar mass ($\sim 1$ dex), but a much wider range in $\Sigma_{5}$ ($\sim 3$ dex). There is also a (weak) correlation of stellar mass and $\Sigma_{5}$, and it is important to account for this by making comparisons to well-defined control samples.

\subsection{The control sample for fifth nearest neighbour analysis}

In order to only examine the connection between radio emission and environment, we construct control samples of galaxies without detected radio emission.  These are matched to have the same mass and colour as the radio galaxies.  We then carry out identical analyses on both the radio galaxies and controls. 

We randomly matched each radio galaxy of a particular class to five\footnote{Our results did not change when we allowed HERGs and LERGs with a minimum of two controls to remain in the analysis.}  GAMA non-radio galaxies (i.e. we exclude all radio galaxies from the control sample) with similar values of stellar mass ($|\Delta\log M_{\star}| < 0.1$ dex) and rest-frame colour ($|\Delta(g-i)_{z=0}| < 0.05$). Each control galaxy is assigned the radio luminosity of the radio galaxy that it is matched to. We remove a control galaxy from the selection pool once it has been matched so that it is not repeated. By matching in stellar mass we remove the weak dependence seen between the stellar mass and $\Sigma_{5}$, while matching in rest-frame colour controls for the known dependence on stellar population with environment \cite[e.g.][]{balogh99}.  We do not match in redshift, as the density defining population is volume--limited. 
We removed radio galaxies for which five control GAMA galaxies could not be found, but minimised the number without five matches by allowing those with the fewest possible controls to match first. The matching process was repeated 100 times to create 100 realisations, and the results were recorded for each realisation. In the analysis we will refer to the median values from the 100 realisations, but the figures that we show are for a single random realisation. Table \ref{tab:den_stats} shows the number of HERGs and LERGs in our final $\Sigma_{5}$ analysis. We distinguish between the radio galaxies and the comparison sample by adding the subscript 'comp' to the class, i.e. X$_{\rm comp}$, where X is HERG or LERG.

\subsection{The local density of radio galaxies}\label{subsec:d5_rg}

The right panel of Fig.\ \ref{fig:denVmass} shows $\Sigma_{5}$ against stellar mass for the LERGs and HERGs with their control galaxies.  Three LERGs were removed because they do not have five non-radio control galaxies (open red circles). Ideally we would like to include these objects, but our aim is to highlight the dependence of environment for radio galaxies beyond the known dependence of stellar mass and stellar population \citep[e.g.][]{balogh99,kauffmann04}.

\begin{table*}
\begin{center}
\caption{Results of our 5th nearest neighbour analysis for the full radio galaxy sample, and the sample divided by radio luminosity (at $L_{\rm NVSS}=10^{23.5}$\,W\,Hz$^{-1}$).  We list the number of galaxies used ($N$), the median 5th nearest neighbour density for the radio galaxies ($\log\Sigma_{5}$) and control galaxies ($\log\Sigma_{\rm 5,comp}$).  We also give the  K--S test probabilities of rejecting the null hypothesis that the radio galaxies are drawn from the same
  distribution as the control.  Lastly, we give $Z(\Delta \rho_{p})$, which is
  the $Z$-score comparing the partial correlation coefficient
  ($\rho_{p}$) of radio galaxies to their control, where $\rho_{p}$ is
  testing the correlation between $\Sigma_{5}$ and radio luminosity
  accounting for redshift dependence.}\label{tab:den_stats}
\begin{tabular}{ll rrrrr}%
\hline
& $L_{\rm NVSS}$ & & Median $\log\Sigma_{\rm 5}$ & Median $\log\Sigma_{\rm5,comp}$\\
Class & {[W Hz$^{-1}$]} & N & [$h^{2} $Mpc$^{-2}$] &  [$h^{2} $Mpc$^{-2}$]  & K--S prob & $Z(\Delta\rho_{p})$\\

\hline
LERG & All & 75 & 0.91 $\pm$ 0.11 & 0.66 $\pm$ 0.05 & 91.5\% &  1.9 $\sigma$\\
LERG &$\leqslant 10^{23.5}$ & 40 & 0.61 $\pm$ 0.15 & 0.56 $\pm$ 0.03& 16.4\% & -\\
LERG &$>10^{23.5}$ & 35 & 1.29 $\pm$ 0.12& 0.79 $\pm$ 0.03& 98.9\% & -\\ 
HERG &All & 13 & --0.07 $\pm$ 0.23  & 0.24 $\pm$ 0.11 & 78.9\% &  1.1 $\sigma$ \\
HERG &$\leqslant 10^{23.5}$ & 7 & --0.21 $\pm$ 0.20 & 0.21 $\pm$ 0.07 & 91.4\% & - \\ 
HERG &$>10^{23.5}$ &  6 & 0.47 $\pm$ 0.36 & 0.31 $\pm$ 0.07 & 32.5\% & - \\
\hline 
\end{tabular}
\end{center}
\end{table*}

The LERGs lie in higher local densities than their control sample, with the median values for the LERGs and controls being $\log\Sigma_5=0.91\pm0.11$ and $0.66\pm0.05$ respectively (see Fig. \ref{fig:den_dn_match} and Table \ref{tab:den_stats}).  The medians are different at the $\simeq2\sigma$ level and a KS-test between the LERGs and their controls shows a marginal (91.5 percent significant) difference in $\Sigma_5$.

We then split the sample into two on radio luminosity, at $\log(L_{\rm NVSS}/{\rm W\,Hz}^{-1}) =23.5$, which is the approximate median luminosity of the sample.  The brighter LERGs have a median $\Sigma_5$ which is $\simeq0.5$\,dex higher than their control sample (significant at $\simeq4\sigma$).  By contrast, fainter LERGs show no significant difference to their controls. 

The HERGs do not lie in significantly different environments compared to their controls (see Fig. \ref{fig:den_dn_match} and Table \ref{tab:den_stats}).  This is largely due to the small number of HERGs  in the sample (only 13 objects), so we are unable to draw any strong conclusions regarding the HERGs at this point.

\section{Radio galaxies in friends-of-friends groups} \label{sec:fof}

We also take a second approach to defining environment, using the GAMA galaxy groups catalogue \citep{robotham11}.  The group catalogue was built using a Friends-of-Friends (FoF) linking of galaxies within a linear projected separation and radial separation defined by mock catalogues.  The original GAMA group catalogue, using just GAMA I data, is described by  \cite{robotham11}.  However, in our work we will use a deeper version of the group catalogue using the GAMA II sample that is complete to $r_{pet}<19.8$ mag in all three equatorial GAMA fields (GAMA Galaxy Group Catalogue version 6, G3Cv6).  The deeper group catalogue allows us to probe lower-mass haloes and increases the multiplicity of existing groups, and hence group properties (e.g. halo mass) are more robust.

\subsection{The control sample for group analysis}\label{subsec:fof_control}

We construct a control sample by matching each radio galaxy to five non-radio GAMA galaxies that were used in the generation of G3Cv6 and have redshift $0.01<z<0.4$. The galaxies were again matched in stellar mass ($|\Delta\log M_{\star}| < 0.1$ dex) and rest-frame colour ($|\Delta(g-i)_{z=0}|< 0.05$), but with an additional constraint on the redshift ($|\Delta z| < 0.01$) due to the redshift dependence of the group completeness (Fig. \ref{fig:fsgama_groupcut}).  The group luminosity,  $L_{\rm FoF}$, is an estimate of the total (rest-frame) $r$-band group luminosity and is calculated for each group by scaling the total observed luminosity with a correction factor to account for undetected faint companions \cite[see][for further details]{robotham11}.  As redshift increases, low luminosity/mass groups drop out of the group sample.  As before, each control galaxy is assigned the radio luminosity of its matched radio galaxy. Radio galaxies for which we could not find five control galaxies (5 HERGs and 45 LERGs) were not used\footnote{Once again, our results did not change when we allowed HERGs and LERGs that had a minimum of two controls to remain in the analysis.}. The number of radio galaxies that satisfied our selection criteria and their median stellar mass are shown in Table \ref{tab:f_group}. 

\subsection{The fraction of radio galaxies in groups}\label{subsec:radio_grps}

\begin{table*}
\begin{center}
\caption{Fraction ($f$) and statistical analysis of radio galaxies and
  their control for different classes (LERG and HERG) that are in
  groups. Testing the significance of the difference between the radio
  galaxies and their control is done using a Bayesian method $P_{N}(D|M_{d})$, which is
  the probability (expressed as a percentage) of our data ($D$) given a model assuming the rates
  are different ($M_{d}$).}\label{tab:f_group}
\begin{tabular}{ll r rrrrr}
\hline
& $L_{\rm NVSS}$ & & Median\\
Class & {[W Hz$^{-1}$]} & N & $\log M_{\star}$ &  $f_{\rm group}$ &$f_{\rm group,comp}$ & $P_{N}(D|M_{d})$\\
\hline
LERG & All & 298 & 10.9 & $0.82^{+0.04}_{-0.05}$ & $0.63^{+0.02}_{-0.03}$ &  $>99.9$\%   \\
LERG &$\leqslant 10^{24}$ & 134 & 10.8 & $0.82^{+0.06}_{-0.07}$ & $0.70^{+0.03}_{-0.04}$ &  86.8\% \\
LERG &$>10^{24}$ & 164 & 11.0 & $0.82^{+0.05}_{-0.07}$ & $0.58^{+0.03}_{-0.03}$ & $>99.9$\% \\
HERG &All & 31 & 10.5 & $0.55^{+0.16}_{-0.17}$ & $0.53^{+0.08}_{-0.08}$ &  19.9\%\\
HERG &$\leqslant 10^{24}$ &13 & 10.2 & $0.69^{+0.18}_{-0.27}$ & $0.55^{+0.12}_{-0.12}$ &  32.2\% \\
HERG &$>10^{24}$ &   18 & 10.6 & $0.44^{+0.22}_{-0.20}$ & $0.49^{+0.10}_{-0.10}$ &  25.3\% \\
\hline 
\end{tabular}
\end{center}
\end{table*}

\begin{figure}
\centering
\includegraphics[width=0.5\textwidth,trim=50 0 0 0]{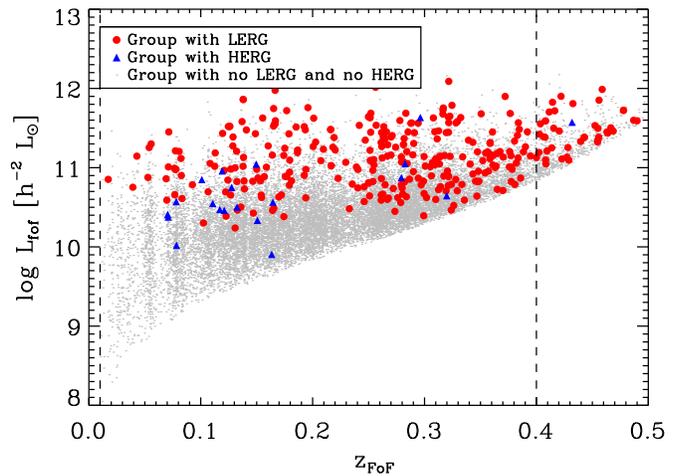}
\caption[GAMA group luminosities as a function of group median redshifts]{GAMA group luminosities ($L_{\rm FoF}$) as a function of group median redshifts ($z_{\rm FoF}$). Grey dots show the full set of GAMA groups, red dots indicate groups that contain a LERG and blue dots are groups with a HERG. The vertical dashed line indicate the redshift limits ($0.01<z<0.4$) imposed for our analysis.}
\label{fig:fsgama_groupcut}
\end{figure}

LERGs tend to lie in the most luminous groups, as can be seen in Fig. \ref{fig:fsgama_groupcut}.  By contrast, the HERGs are in lower luminosity groups.  However, from our analysis above (e.g. see Fig. \ref{fig:denVmass}), we know that HERGs also have lower stellar mass.  This difference in stellar mass between HERGs and LERGs could be driving the difference in group luminosity. 

We find that LERGs are significantly more likely ($P_{N}(D|M_{d}) > 99.9$ percent, $Z = 3.5\sigma$, see Table \ref{tab:f_group}) to lie in galaxy groups ($82$ percent are in groups) when compared to a matched radio-quiet control sample ($\sim63$ percent are in groups).  We again subdivide the sample into two, based on luminosity.  In this case we make the division at $\log(L_{\rm NVSS}/{\rm W\,Hz}^{-1}) =24$, which is the approximate median of the group sample (this is higher than the luminosity of the $\Sigma_5$ sample due to the broader redshift range of the groups).  We find that only the high-luminosity LERGs (see Table \ref{tab:f_group}) show a significant difference compared to their control sample, with the fraction in a group being $0.82^{+0.05}_{-0.07}$ and $0.58^{+0.03}_{-0.03}$ for radio galaxies and controls respectively. This result is qualitatively consistent with the difference in local density seen in \S\ref{sec:d5}, again implying that some feature of the group environment acts to enhance the observed radio emission in low-excitation radio galaxies.  

  {\update The importance of controlling for redshift in our group analysis is highlighted by the fact that the control galaxies for the high luminosity LERGs have a lower group fraction than the controls for the low luminosity LERGs, even though the high luminosity LERGs have higher stellar mass.  This is driven by the high luminosity LERGs typically having higher redshift, where only the more massive groups are detected.}  

We see no difference between the fraction of HERGs and controls in groups, including if we sub-divide by radio luminosity, even though we have a factor of 2 more HERGs than in the $\Sigma_{5}$ sample analysed in \S\ref{sec:d5}.  However, the small number of HERGs means that this is still not a strong constraint.  A difference between the fraction of radio galaxies and controls in groups equal to the result we find for LERGs would only be detectable at the $\simeq1\sigma$ level in the HERGs.  

\section{The properties of groups hosting radio galaxies}\label{sec:fof_prop}

We now investigate the properties of the groups that radio galaxies reside in. We consider the fraction of radio galaxies that are the brightest group member; the fraction of radio galaxies that are the centre galaxy in a group; the linear projected distance between each radio galaxy and the group centre; the dynamical mass of the groups ($M_{\rm FoF}$); and the distance between radio galaxies and their nearest group member.

Now that we are considering the relative difference of radio galaxies to other galaxies in groups, we need to remake our control samples.  This time, in addition to the constraints on stellar mass, colour and redshift, we include the further constraint that the control galaxies must also be in a group.  All other constraints were the same as in \S\ref{subsec:radio_grps}. The corresponding numbers after the rematching are listed in Table \ref{tab:bcg_stats}.

\subsection{The fraction of radio galaxies that are the brightest or central galaxy in a group}\label{subsec:bcg}

The brightest cluster or group galaxy (BCG\footnote{Even though many of the GAMA groups have halo masses below the typical halo mass of a cluster \citep[$\sim 10^{15} h^{-1} M_{\sun}$;][]{eke96,feretti12}, we will be consistent with \cite{robotham11} and refer to the brightest group/cluster member as the BCG.}) is typically the dominant (and possibly central) galaxy of any cluster or group.  When we compare the fraction of radio galaxies and controls that are BCGs in a group, we find that there is no significant difference (see Table \ref{tab:bcg_stats}).  The most significant difference is between radio--luminous ($L_{\rm NVSS} > 10^{24}$ W Hz$^{-1}$ ) LERGs and their controls, but this is only significant at the 90 percent level.  

Alternatively to the BCGs, we can examine whether being the galaxy closest to the group centre influences radio properties.  The BCG will generally be the most massive galaxy, but it need not be the galaxy in the deepest part of the group potential, although this is more likely when the group is virialized.  The central galaxy in each group was determined through an iterative process, where the most distant galaxy from each iteration of the $r$-band centre-of-light calculation, was rejected. The final galaxy that remained in this process became the iterative central galaxy (IterCen) of the group. \cite{robotham11} showed that the IterCen provided the best match to the group centres in the mock catalogue, therefore we consider the IterCen as the central galaxy of a group. For low-multiplicity groups however, the group centres may not be well defined (the median membership is about 4 for the radio galaxies and their control samples). 

As shown in Table \ref{tab:bcg_stats}, we find that there is no significant difference between the likelihood of a radio galaxy or control galaxy being the central galaxy in a group.  This suggests that once mass, colour and redshift are accounted for, there is not a strong preference for radio galaxies to be the most massive or central galaxy in a group.


The fraction of central galaxies estimated here for high mass galaxies (both LERGs and their control sample) is lower than that from halo occupation distribution (HOD) models for galaxies of a similar mass.  The typical median stellar mass for our LERGs is $\simeq10^{11}$\,M$_{\odot}$.  This is equivalent to an absolute magnitude in the $r$-band of $M_r\simeq-21.6$, assuming a mass-to-light ratio of $\log(M_*/L_r)=0.5$ \citep{kauffmann03}.  \cite{2011ApJ...736...59Z} find a satellite fraction of $f_{\rm sat}=0.15\pm0.01$ for galaxies brighter than $M_r\simeq-21.0$ (and $f_{\rm sat}=0.09\pm0.01$ for galaxies brighter than $M_r\simeq-21.5$).  However, the fitted satellite fraction is a strong function of colour, with Zehavi et al.\ finding that at $-21<M_r<-20$ the satellite fractions for red--sequence and blue--cloud galaxies are $\simeq0.3$ and 0.13 respectively.  The LERG population lies almost exclusively on the red sequence so will have a higher satellite fraction than a purely stellar mass limited sample.  

The mass threshold for detecting a group also biases the central fraction to be lower, as massive galaxies are more likely to be the central galaxy if they inhabit a lower mass group.  However, such biases do not influence our differential tests, as our controls are selected to have the same mass, colour and redshift as the radio-galaxies.

\begin{table*}
\begin{center}
\caption{Fraction ($f$) and statistical analysis of radio galaxies and
  their control sample that are in a group and are either the BCG or
  the central galaxy (IterCen). 
  $P_{N}(D|M_{d})$ is the probability that the data fits a model in
  which the rates are different.}\label{tab:bcg_stats}
\begin{tabular}{l r rr rrrr}%
\hline
$L_{\rm NVSS}$ & \multicolumn{2}{c}{$f({\rm BCG})$} &&& \multicolumn{2}{c}{$f({\rm BCG})$} \\
{[W Hz$^{-1}$]} & LERG & LERG$_{\rm comp}$ & $P_{N}(D|M_{d})$ & HERG & HERG$_{\rm comp}$ & $P_{N}(D|M_{d})$  \\
\hline\hline
All & $0.742^{+0.052}_{-0.060}$ & $0.651^{+0.027}_{-0.028}$ &  74.3\% & $0.625^{+0.191}_{-0.242}$ & $0.600^{+0.101}_{-0.110}$ & 24.9\% \\
$\leqslant 10^{24}$ & $0.713^{+0.077}_{-0.092}$ & $0.681^{+0.038}_{-0.041}$ &  12.7\% & $0.750^{+0.175}_{-0.350}$ & $0.550^{+0.143}_{-0.153}$ & 38.2\% \\
$>10^{24}$ & $0.766^{+0.065}_{-0.081}$ & $0.625^{+0.037}_{-0.038}$ & 90.4\% & $0.500^{+0.288}_{-0.288}$ & $0.650^{+0.129}_{-0.156}$ & 35.2\% \\
\hline
$L_{\rm NVSS}$ & \multicolumn{2}{c}{$f({\rm IterCen})$} &&& \multicolumn{2}{c}{$f({\rm IterCen})$} \\
{[W Hz$^{-1}$]} & LERG & LERG$_{\rm comp}$ & $P_{N}(D|M_{d})$ & HERG & HERG$_{\rm comp}$ & $P_{N}(D|M_{d})$ \\
\hline\hline
All & $0.713^{+0.054}_{-0.061}$ & $0.641^{+0.027}_{-0.028}$ & 40.9\% & $0.688^{+0.170}_{-0.247}$ & $0.613^{+0.099}_{-0.110}$ & 26.5\% \\
$\leqslant 10^{24}$ & $0.713^{+0.077}_{-0.092}$ & $0.681^{+0.038}_{-0.041}$ & 12.8\% & $0.750^{+0.175}_{-0.350}$ & $0.550^{+0.143}_{-0.153}$ &  41.2\% \\
$>10^{24}$ & $0.711^{+0.071}_{-0.084}$ & $0.608^{+0.037}_{-0.038}$ & 53.7\% & $0.625^{+0.238}_{-0.326}$ & $0.650^{+0.129}_{-0.156}$ & 30.8\% \\
\hline 
\end{tabular}
\end{center}
\end{table*}

\subsection{The fraction of BCGs that are radio galaxies}

In the previous section we asked the question, what is the fraction of radio galaxies in a group that are the brightest or central galaxy in that group?
We can also ask the reverse question; given that a galaxy is a BCG, what is the chance that it is a radio galaxy?  Previous studies \citep[e.g.][]{burns81,menon85} have shown that BCGs in high mass clusters can often host a radio-loud AGN.  \cite{best07} studied a sample of 625 nearby groups and clusters and showed that BCGs were more likely to host a radio-loud AGN than other galaxies (including those that are non-BCG cluster galaxies).  The difference is a factor of ten for galaxies with stellar mass below $10^{11} M_{\sun}$, but less than a factor of two for stellar masses of $\sim5\times10^{11} M_{\sun}$.  The Best et al.\ analysis used groups with a median velocity dispersion of $\sigma_{\rm group}\simeq400$ km s$^{-1}$ at $z<0.1$ \citep{miller05,von-der-linden07}. By comparison the groups in our sample have a median dispersion of $\sigma_{\rm group}=170$ km s$^{-1}$.

\begin{figure}
\centering
\includegraphics[width=0.5\textwidth, trim = 50 0 0 0]{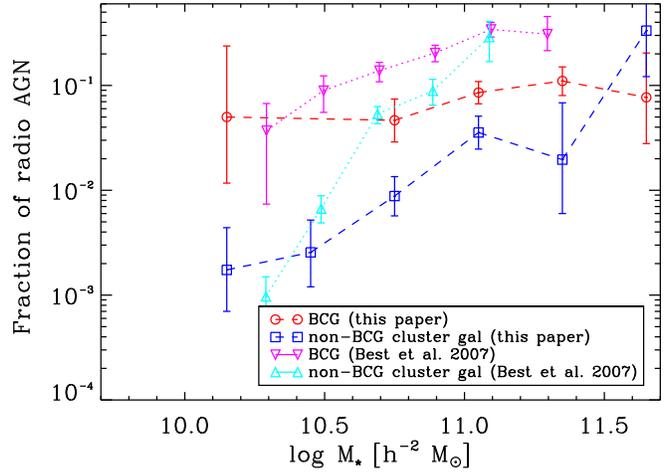}
\caption[The fraction of BCGs and non-BCG cluster galaxies that host a radio AGN as a function of stellar mass]{The fraction of BCGs (red circles) and non-BCG cluster galaxies (blue squares) that host a radio AGN (regardless of high- or low- excitation) as a function of stellar mass. We have defined clusters as groups with group luminosity brighter than $10^{11} h^{-2} L_{\sun}$ and $z_{\rm FoF}<0.4$. The results from the \citet{best07} work for BCGs are in magenta inverted triangles and non-BCG cluster galaxies are in cyan triangles.}
\label{fig:pbest07}
\end{figure}

We construct a set of volume-limited groups by only using groups with $L_{\rm FoF} > 10^{11} h^{-2} L_{\sun}$ and $z_{\rm FoF}<0.4$ and calculate the fraction of BCGs and non-BCG group galaxies that host a radio AGN as a function of stellar mass (see Fig. \ref{fig:pbest07}).  In order to directly compare to the work of Best et al.\ (2007) in this analysis we combine both HERGs and LERGs into one sample that we simply call radio AGN {\update (although we note that the numbers and trend are dominated by LERGs)}.

At low stellar masses, our BCGs have a higher probability of hosting a radio AGN than non-BCG cluster galaxies. The probability of the data, given a model where the two fractions are drawn from different distribution, is $P_{N}(D|M_{d})>99\%$ and $=99\%$ for the $10^{10.75}$ and $10^{11.05}$ stellar mass bins respectively, and drops below 95\% for higher stellar mass bins.   This is in qualitative agreement with the results of \cite{best07}, but the fraction of galaxies that are radio AGN in our sample is lower than that found by Best et al.  This is not surprising as our sample has a similar radio flux limit to that of Best et al., but a higher median redshift, such that the radio detection rate is on average lower.

\subsection{Distance from the group centre}\label{subsec:R_dn}

\begin{figure*}
\centering
\includegraphics[width=90mm,trim=20 0 20 0]{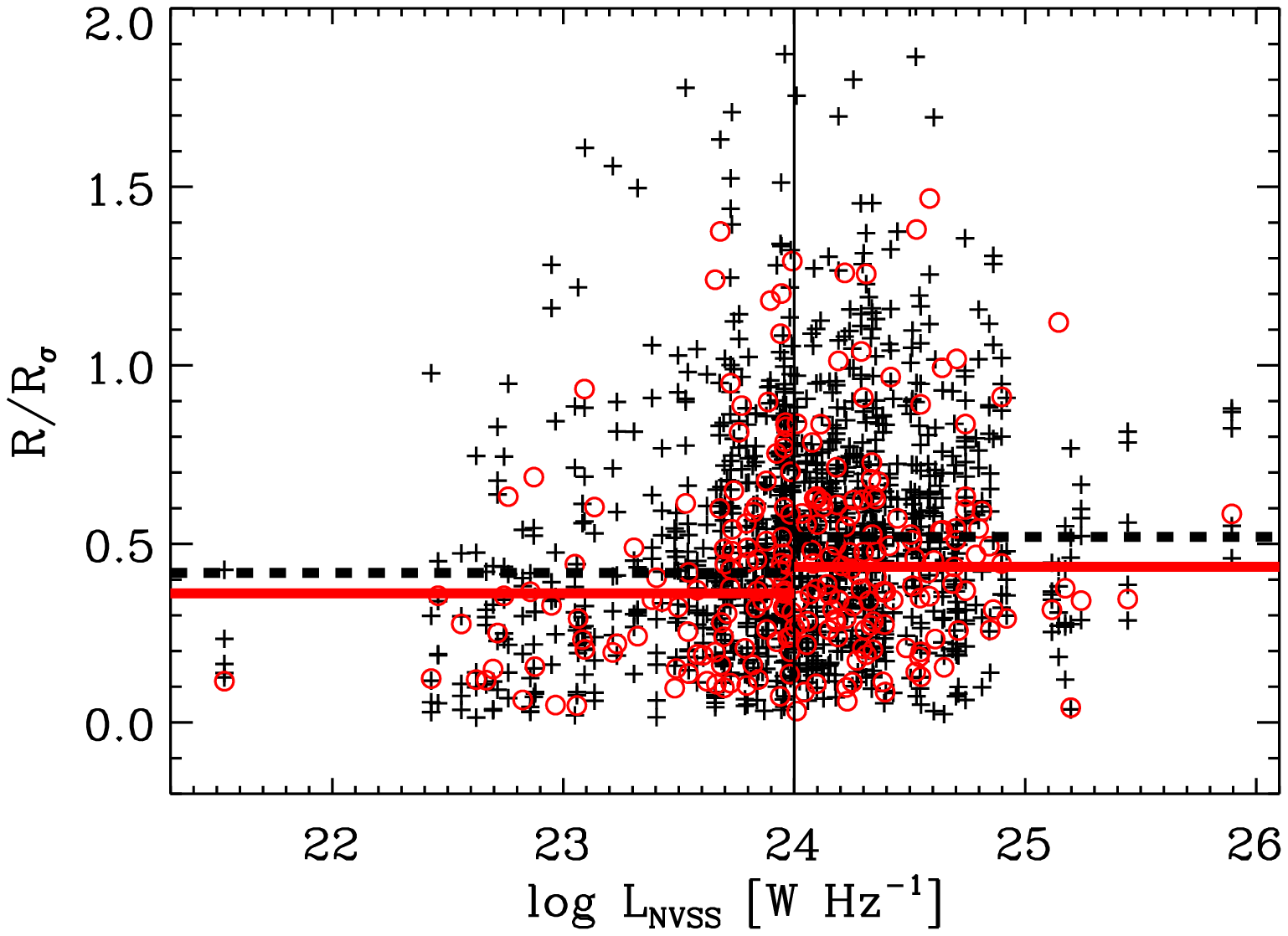}\includegraphics[width=90mm,trim=20 0 20 0]{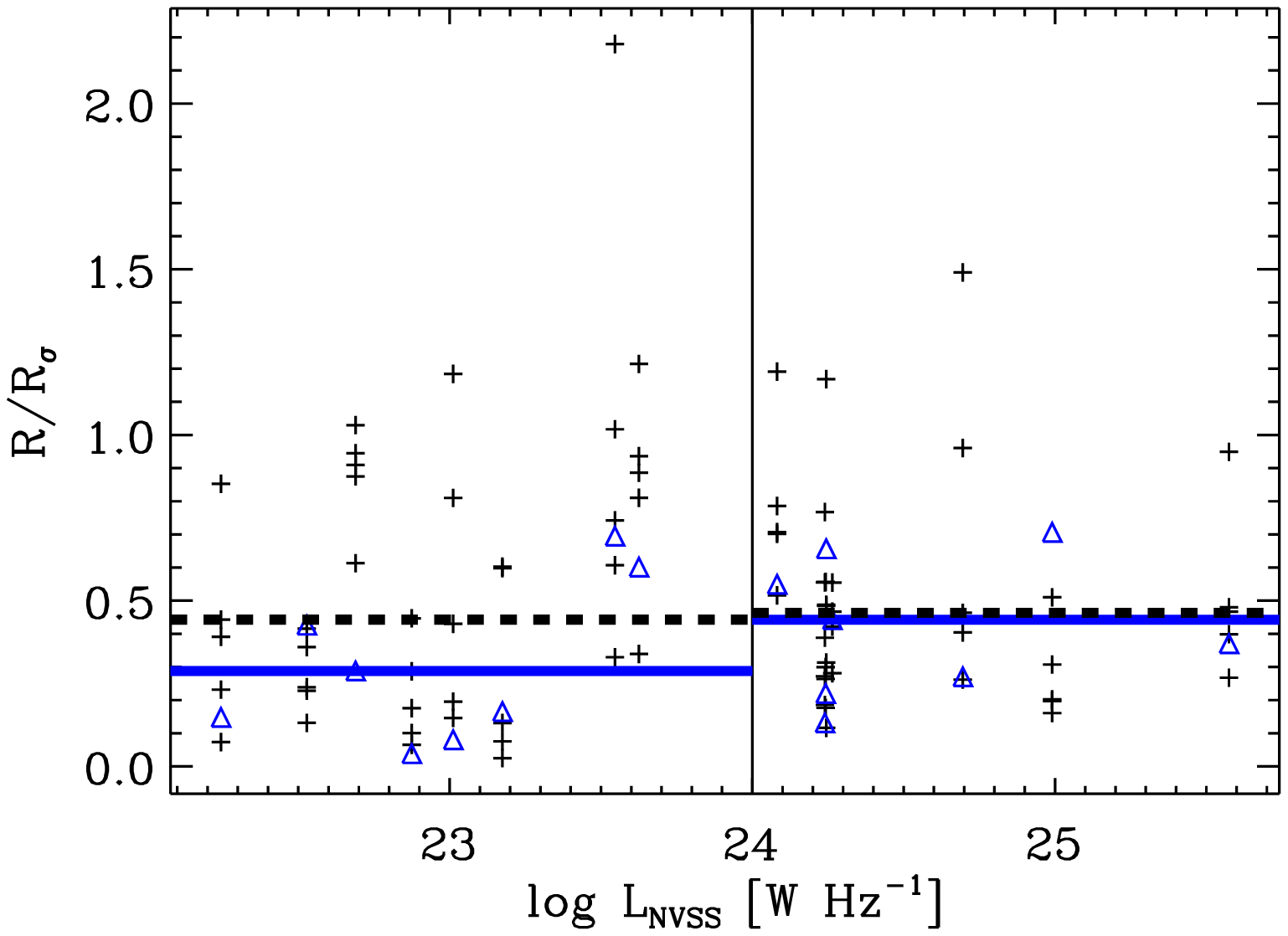}
\caption{The normalized radial distance of radio galaxies from the group centre as a function of radio luminosity.  Left: LERGs (red circles) and their controls (black crosses); right: HERGs (blue triangles) and their controls (black crosses).  The radial distance is normalized by the radius containing 68 percent of the group members, $R_{\sigma}$ for each group.  The horizontal lines indicate the median value of the radial distribution for radio galaxies (solid lines) and their control samples (dashed black lines) in the two different radio luminosity bins.}
\label{fig:Rgrp_cumdn}
\end{figure*}

Next we examine the radial position of radio galaxies within the GAMA groups and find that LERGs preferentially lie in the inner regions of the groups compared to their control samples (see Fig.\ \ref{fig:Rgrp_cumdn}).  We use the centre-of-light positions \cite[see][]{robotham11} that are not associated with a specific galaxy and measure the projected radial distance of  a galaxy, $R$, in units of the radius containing 68 percent of the group members, $R_{\sigma}$).  LERGs are preferentially found at smaller radius from the group centre than their controls.  A two sample K-S test (Table \ref{tab:ks_rdn}) rejects the null hypothesis that the LERGs and controls are draw from the same sample at  the 99.7\% level.   When we divide the sample in two by luminosity, only the high luminosity ($L_{\rm NVSS} \leqslant 10^{24}$ W Hz$^{-1}$) LERGs show a significant difference from their controls in radial distribution.

\begin{table*}
\begin{center}
\caption{KS-test probability of rejecting the null hypothesis that the projected radial distribution of radio galaxies in groups is the same as their controls, for: the full sample, non-BCG group galaxies, non-IterCen group galaxies, BCGs (only for LERGs), IterCen galaxies (only for LERGs).  Results are shown separately for the HERG$^{\rm FIRST}$ sample (including all LERGs to the FIRST detection limit).  The number in parenthesis after the probability is the number of  radio galaxies considered in each analysis. The last column is the $Z$-score comparing the Spearman rank correlation coefficient ($\rho_{s}$) for radio galaxies and their control sample, where we test the correlation between radio luminosity and radial distribution. The control galaxies adopt the radio luminosity of their matched radio galaxy.}
\label{tab:ks_rdn}
\begin{tabular}{l r rrr }%
\hline
 & \multicolumn{3}{c}{$L_{\rm NVSS}$ [W Hz$^{-1}$]}\\
Class & All & $\leqslant 10^{24}$ & $>10^{24}$& $Z(\Delta\rho_{s}$)\\
\hline\hline
LERG (all group galaxies)& 99.7\% (236) & 82.6\% (108) & 99.8\% (128) &  0.2$\sigma$\\
HERG (all group galaxies) & 67.3\% (16) & 68.5\% (8) & 16.0\% (8) &  1.3$\sigma$\\
LERG (non-BCG group galaxies) & 80.2\% (53) & 17.5\% (29) & 90.4\%\ (24) & 1.0$\sigma$\\
HERG (non-BCG group galaxies) & 96.9\% (6) & - (2) & - (4) & 0.4$\sigma$\\
LERG (Non-IterCen group galaxies) & 71.3\% (63) & 7.6\% (30) & 89.2\% (33) & 0.4$\sigma$\\
HERG (Non-IterCen group galaxies) & 98.8\% (5) & - (2) & - (3) & 0.3$\sigma$\\
LERG (BCGs only) & 90.1\% (165) & 67.2\% (74) & 65.7\% (91) &  1.1$\sigma$\\
LERG (IterCen galaxies only) & 96.3\% (156) & 85.0\% (73) & 72.1\% (84) &  1.0$\sigma$\\
HERG$^{\rm FIRST}$ (All group galaxies) & 95.6\% (53) & 99.2\% (25) & 30.6\% (28)&  2.4$\sigma$\\
HERG$^{\rm FIRST}$ (Non-BCG group galaxies) & 92.4\% (21) & 94.9\% (9) & 41.3\% (12) &  0.3$\sigma$\\
HERG$^{\rm FIRST}$ (Non-IterCen group galaxies) & 90.6\% (21) & 94.9\% (9) & 32.0\% (12) &  0.3$\sigma$\\
\hline 
\end{tabular}
\end{center}
\end{table*}

The relationship between radio emission and distance from group centre is dominated by the contribution of galaxies that are BCGs.  We repeat our radial analysis with the BCG and the IterCen galaxies removed, including also remaking our control samples so that non-radio BCGs and IterCen galaxies are removed. The radial distribution of LERGs no longer shows a significant difference to the control sample once the BCGs and IterCen galaxies are removed.  This is not simply caused by the reduced number of LERGs because applying a similar analysis to 25 random LERGs from the full sample (and using their associated control sample) still results in a $\sim99\%$ significant difference for the radial distribution. Thus, the difference in the radial distribution for all LERGs in groups is largely influenced by the subset that are the BCG or the central galaxy of their group.

If we consider only BCGs or central galaxies and test whether there is any preference for those with radio emission to be closer to the group centre, we find a marginally significant difference between the radio galaxies and their controls, at the 96 percent and 90 percent significance level for the IterCen galaxies and BCGs respectively.  This suggests a plausible hypothesis that both being the BCG or central galaxy {\it and} being close to the centre of the group potential influences radio emission.
 
\begin{figure}
\centering
\includegraphics[width=90mm,trim= 50 0 0 0]{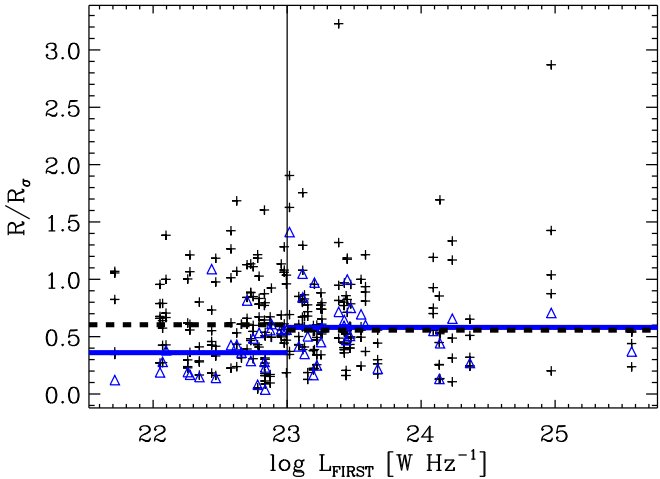}
\caption[]{The projected radial separation between HERGs/controls (triangles/crosses) and group centres as a function of radio luminosity for the larger HERG$^{\rm FIRST}$ population which only requires a FIRST detection. Horizontal lines denote the median value of radial distribution for radio galaxies (solid) and their controls (dashed).}
\label{fig:Rgrp_cumdn_FIRST}
\end{figure}

HERGs show no difference in their radial distribution with respect to their control sample, but the sample is small (see Fig. \ref{fig:Rgrp_cumdn}, right).  To more robustly examine trends in the HERG sample we require a larger sample, as HERGs are intrinsically rarer than LERGs (at least within the redshift range we are probing).  To expand the HERG sample we augment them with objects that were detected in FIRST ($\simeq1$\,mJy limit), but fall below the flux limits applied earlier ($S_{\rm tot}>3.5$ mJy, including the requirement to be detected in NVSS).  We denote this revised sample as HERG$^{\rm FIRST}$.  Reapplying our radial distribution analysis to the larger HERG$^{\rm FIRST}$ sample find that HERGs are marginally closer to the centre of their group than their control sample (95 percent significant, see Table \ref{tab:ks_rdn} and Fig.\ \ref{fig:Rgrp_cumdn_FIRST}).  The difference becomes stronger for the low-luminosity ($L_{\rm FIRST} <10^{24}$\,W\,Hz$^{-1}$) HERGs, at the 99 percent level.  Comparison of the Spearman rank correlation coefficients for HERGs and their control sample shows that the dependence of radial distribution on radio luminosity is significant to $2.6\sigma$.  Removing the BCGs and IterCen galaxies reduces the significance of the difference for both the full sample and the low-luminosity sub-sample.

\subsection{Group mass}\label{subsec:grp_mass}

Cross-correlation studies of radio galaxies \citep[e.g.][]{wake08,mandelbaum09,donoso10,2014MNRAS.440.1527L,2014MNRAS.440.2322L} have shown that LERGs tend to lie in more massive haloes than radio-quiet galaxies with similar intrinsic properties (e.g. stellar mass, luminosity, redshift).

For a virialized system, the dynamical mass can be estimated as $M\propto r\sigma^{2}$, where $r$ and $\sigma$ are the radius and velocity dispersion of the group. G3Cv6 group masses were estimated by assuming the group radius to be the radius containing 50\% of the group ($R_{50}$) and adopting the group velocity dispersion ($\sigma_{\rm FoF}$) estimated using the \cite{beers90} method because it is less sensitive to changes in group membership ($N_{\rm FoF}$). An additional scaling factor $A(N_{\rm FoF}, z_{\rm FoF})$ was used to calibrate the mass estimates to mock groups from simulations.  Thus the dynamical mass estimates have the form $M_{\rm FoF} = A(N_{\rm FoF}, z_{\rm FoF}) R_{50} \sigma_{\rm FoF}^2$.

\begin{figure*}
\centering
\includegraphics[width=90mm,trim=20 0 20 0]{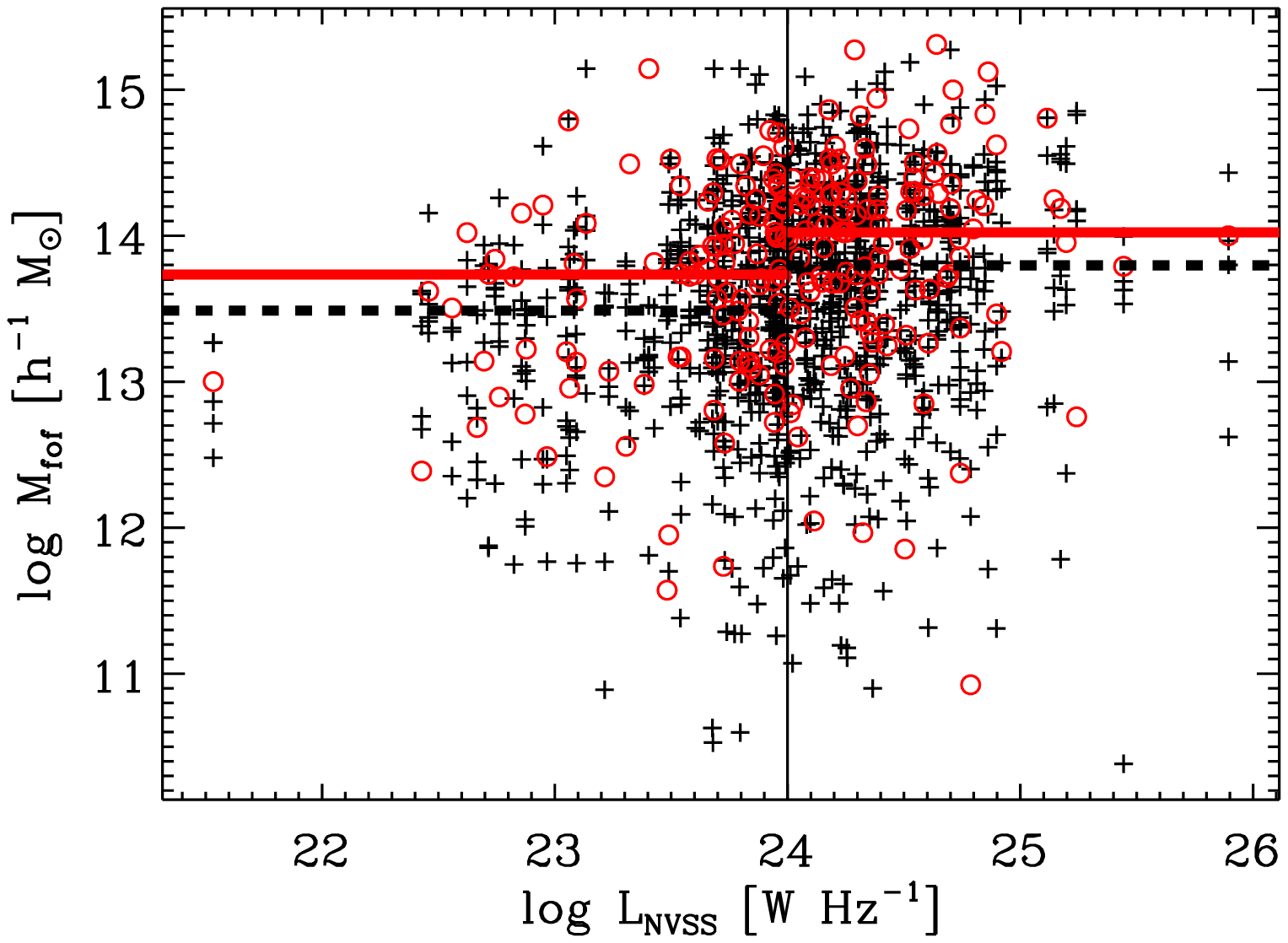}\includegraphics[width=90mm,trim=20 0 20 0]{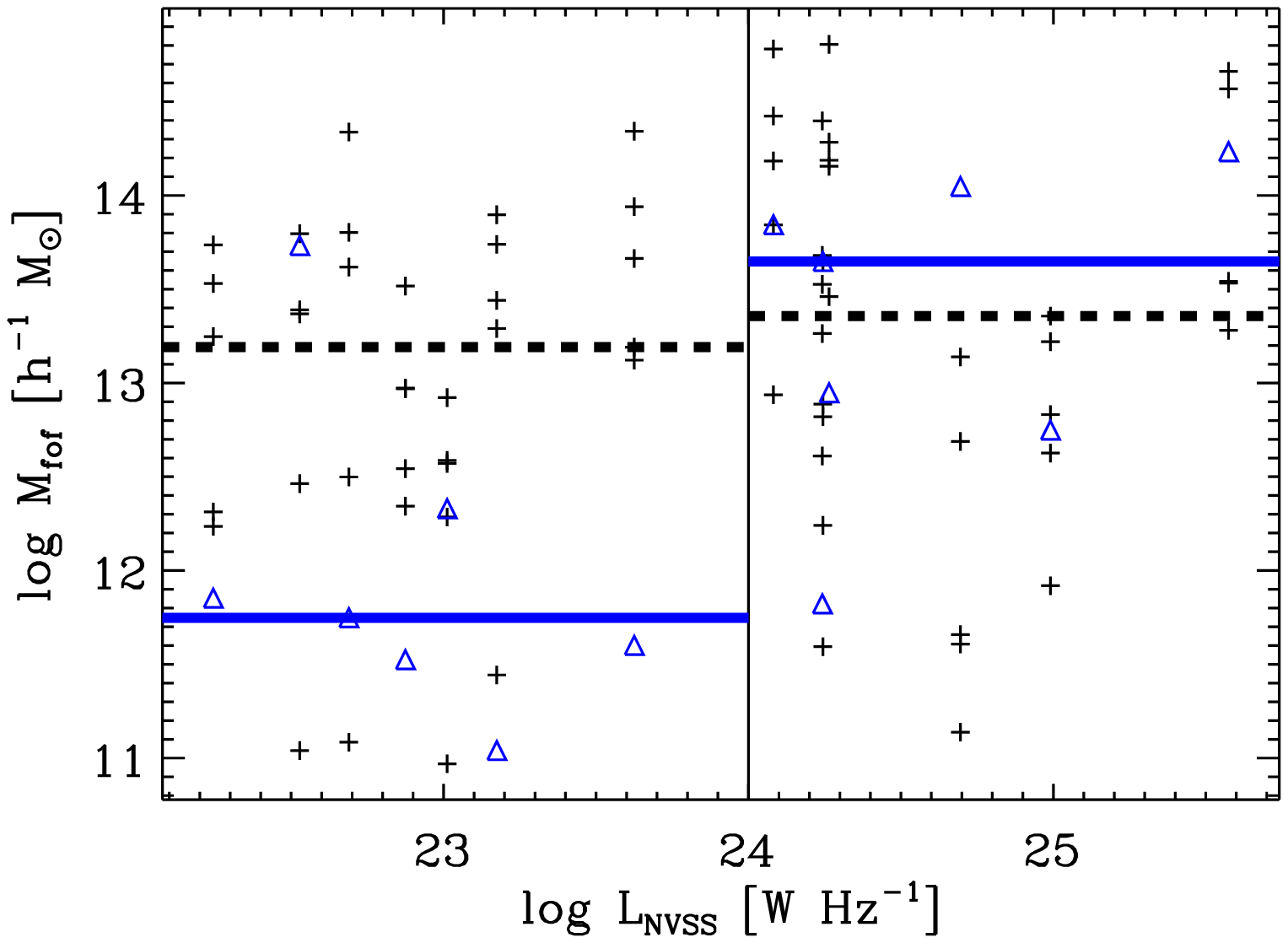}
\caption[The group dynamical mass as a function of radio luminosity for groups that contain a radio galaxy and their control sample]{The group dynamical mass ($M_{\rm FoF}$) as a function of radio luminosity, for groups that contain a radio galaxy (LERGs; red circles in the left panel, and HERGs; blue triangles in the right panel) and their control sample (crosses). The horizontal lines indicate the median values of $M_{\rm FoF}$ (solid and dashed lines for radio galaxies and their control sample respectively) for $L_{\rm NVSS} \leqslant 10^{24}$ W Hz$^{-1}$ and $L_{\rm NVSS} > 10^{24}$ W Hz$^{-1}$.}
\label{fig:mfof}
\end{figure*}

Fig.\ \ref{fig:mfof} shows the $M_{\rm FoF}$ as a function $L_{\rm NVSS}$ for radio galaxies and their control samples. The median values of $M_{\rm FoF}$ for each class are also listed in Table \ref{tab:ks_mfof}, along with the two-sample KS-test probability of rejecting the null hypothesis that $M_{\rm FoF}$ for radio galaxies is drawn from the same distribution as their control samples. Both high- and low- luminosity LERGs are in higher ($\sim0.2$ dex) halo masses than their control samples, and the difference is significant at $99.9\%$ for the full sample of LERGs. Similar results are also seen when we compare other proxies for halo mass e.g. $\sigma_{\rm FoF}$ and $L_{\rm FoF}$.

The HERG population inhabits lower mass haloes than their control sample, but the difference is not statistically significant unless we subdivide the sample by luminosity.  Low-luminosity HERGs tend to lie in lower mass haloes than their control sample (99 percent significant), but note that the sample is small with only 7 HERGs in the low-luminosity bin. When considering the expanded HERG$^{\rm FIRST}$ sample the halo mass of HERGs was no longer statistically different to their controls.

\begin{table*}
\begin{center}
\caption{The result of group mass and distance to nearest neighbour analysis.  $N$ is the number of radio galaxies used in each analysis.  We list the median value of group dynamical mass for radio galaxies ($M_{\rm FoF}$) and control galaxies ($M_{\rm FoF,comp}$), together with the K--S test probability that the $M_{\rm FoF}$ of radio galaxies is drawn from the same distribution as the control sample. $Z(\Delta\rho_{s})$ is the $Z$-score when comparing the Spearman rank correlations ($\rho_{s}$) of radio galaxies and their control, where the correlation is between radio luminosity and $M_{\rm FoF}$.  We also include the result of the K--S test comparing the distance to the nearest group member for radio galaxies and the control sample. The distance is defined by either the quadrature sum of the projected and radial distances ($R_{\rm nm}$), or the projected distance only ($R_{\rm nm,proj}$), both in co-moving coordinates.  Finally, we also present the $Z$-score values for the correlation between radio luminosity and nearest neighbour distance.
}
\label{tab:ks_mfof}
\begin{tabular}{ll rrrrrrrrr }%
\hline
& $L_{\rm NVSS}$ &  & $\log (M_{\rm FoF})$ & $\log (M_{\rm FoF,comp})$ & $M_{\rm FoF}$ & $M_{\rm FoF}$ & $R_{\rm nm}$ & $R_{\rm nm}$ & $R_{\rm nm,proj}$ & $R_{\rm nm,proj}$\\
Class &  [W Hz$^{-1}$] & $N$ & [$ h^{-2} M_{\sun}$] & [$ h^{-2} M_{\sun}$] & K--S prob & $Z(\Delta\rho_{s})$& K--S prob & $Z(\Delta\rho_{s})$& K--S prob & $Z(\Delta\rho_{s})$\\
\hline
LERG & All & 226 & 13.9& 13.7& 99.9\% &  0.9$\sigma$ & 91.0\%&  1.0$\sigma$ & 99.9\% &  0.4$\sigma$\\
LERG &$\leqslant 10^{24}$ & 106  & 13.7& 13.6& 96.9\% & - & 96.6\% & -& 99.8\% & - \\
LERG &$>10^{24}$ & 120 & 14.0& 13.8& 98.0\% & - & 14.2\% & - & 92.9\% & - \\
HERG &All & 14 & 12.7 & 13.3 & 92.2\% &  1.3$\sigma$  & 99.4\%&  0.8$\sigma$ & 98.9\% &  1.3$\sigma$ \\
HERG &$\leqslant 10^{24}$ &7  & 11.7& 13.1& 99.6\%& - & 93.3\% & - & 99.4\% & - \\
HERG &$>10^{24}$ &   7 & 13.6& 13.4& 12.3\%& -& 93.3\% & - & 68.5\% & - \\

\hline 
\end{tabular}
\end{center}
\end{table*}

\subsection{Distance to nearest member}\label{subsec:r_nm}

Within each group we find the distance to the nearest group member, to test for evidence that galaxy--galaxy interactions play a role in radio galaxy triggering. For each group galaxy, we find the distance to the nearest group member that it is linked with \citep[see][for linking definitions]{robotham11}.  We use both a projected angular distance (converted to comoving distance based on the mean redshift of the pair), $R_{\rm nm,proj}$, and a three-dimensional distance, $R_{\rm nm}$, that is the quadrature sum of the projected and radial separations in comoving coordinates.  As we are only using galaxies that are already associated with each other via the group catalogue, gross projection effects should be minimised in these nearest neighbour distance estimates.  The three-dimensional distance may be influenced by peculiar motions that will be increasingly important in higher mass groups.  For 28 percent of GAMA group galaxies the $R_{\rm nm,proj}$ and $R_{\rm nm}$ distances give a different nearest neighbour.

\begin{figure*}
\centering
\includegraphics[width=90mm,trim=20 0 20 0]{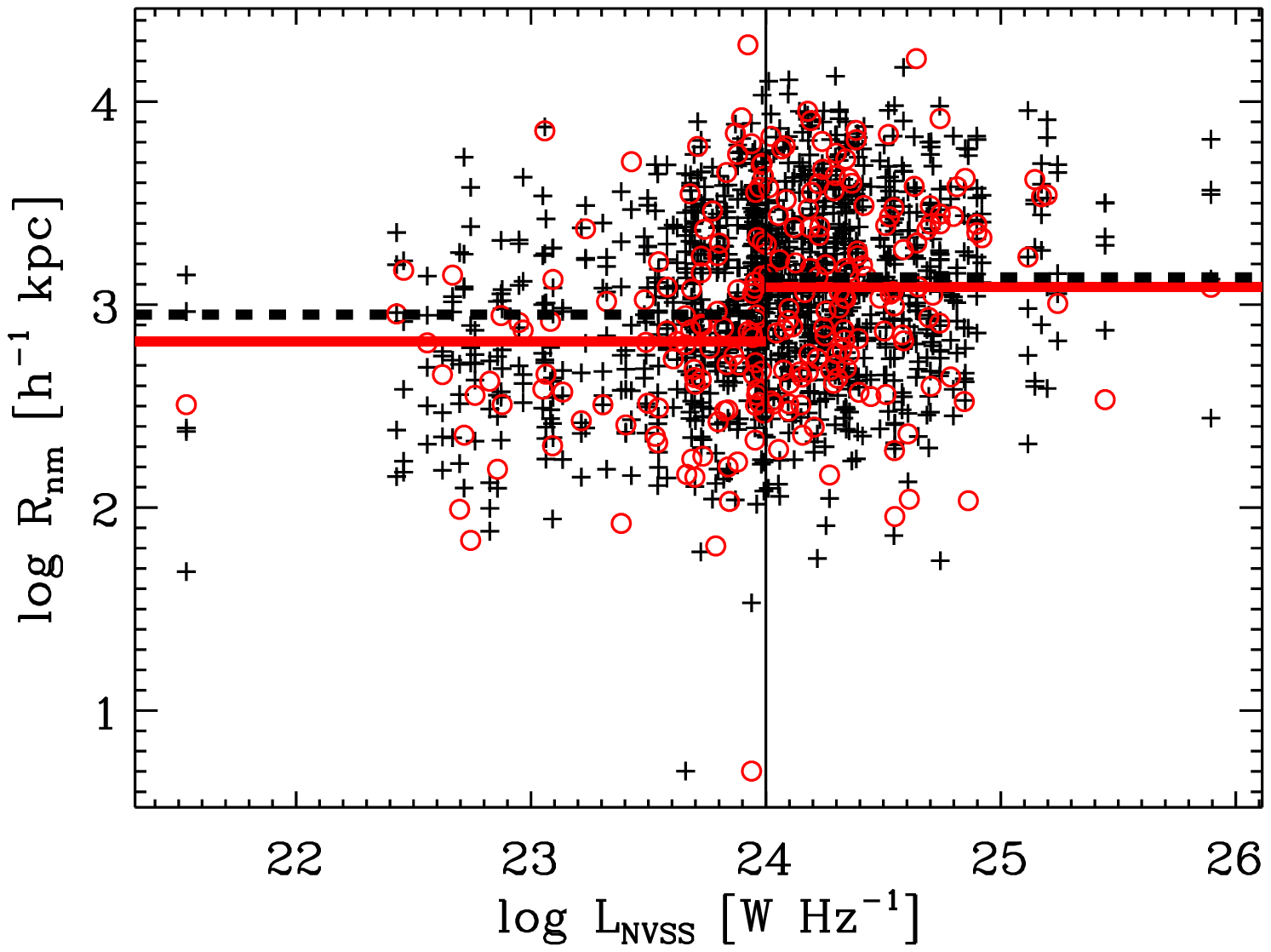}\includegraphics[width=90mm,trim=20 0 20 0]{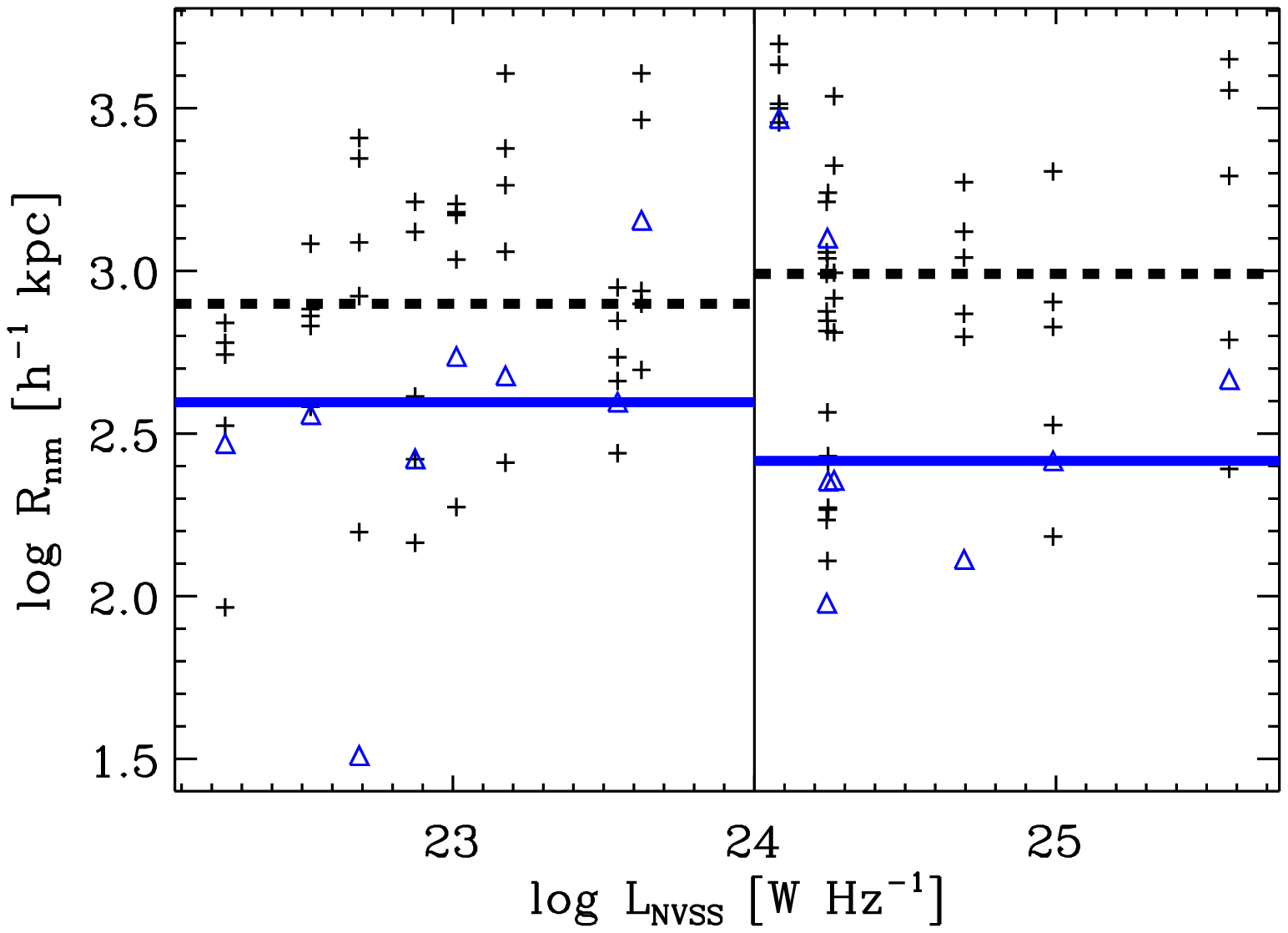}
\includegraphics[width=90mm,trim=20 0 20 0]{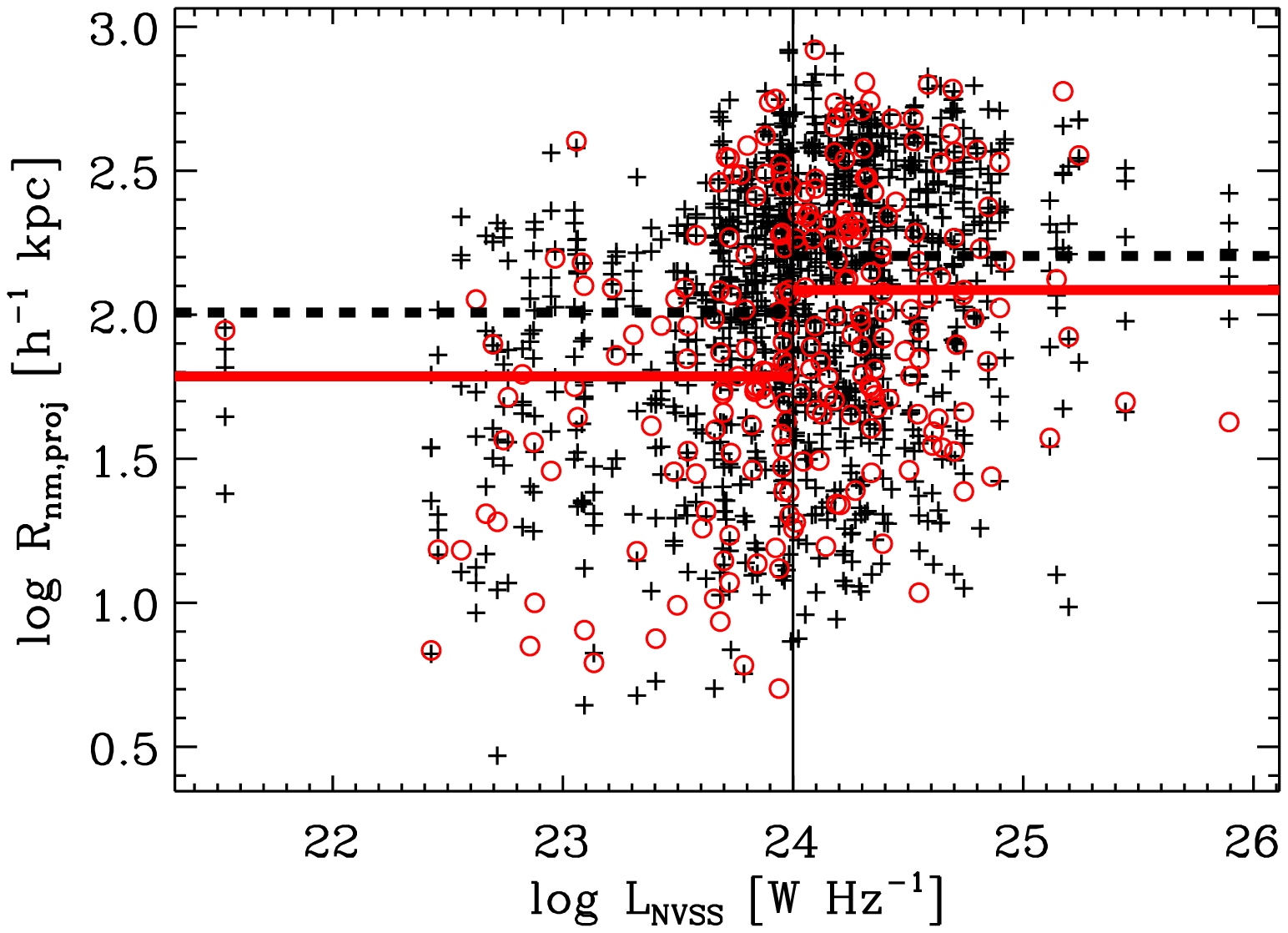}\includegraphics[width=90mm,trim=20 0 20 0]{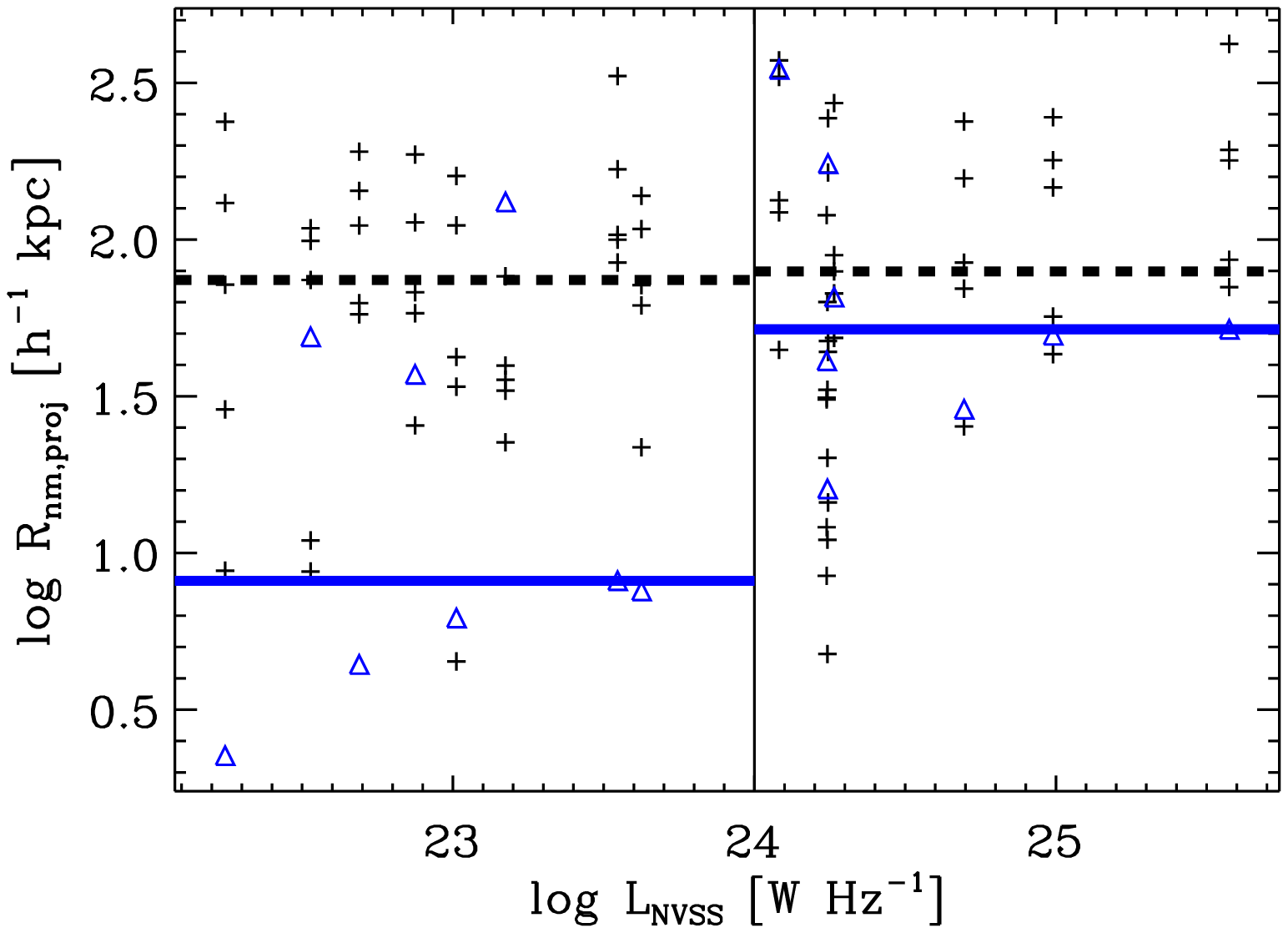}
\caption[Distance from the radio galaxy and their control sample to the nearest group member as a function of radio luminosity]{The distance from the radio galaxies and their control sample (crosses) to the nearest group member as a function of radio luminosity for LERGs (left, red circles) and HERGs (right, blue triangles) separately.  We use two different estimates of nearest neighbour distance; a three dimensional distance (top) and a projected distance (bottom). The horizontal lines indicated the median values of $R_{\rm nm}$ and $R_{\rm nm,proj}$ in high- and low- luminosity bins (solid for radio galaxies and dashed for control galaxies).}
\label{fig:nm}
\end{figure*}

We find that LERGs have significantly closer nearest neighbours than their controls, at the 99 percent significance level (for $R_{\rm nm,proj}$, this drops to 91 percent for $R_{\rm nm}$).  The difference in distance is  $\sim 0.2$ dex (see Fig. \ref{fig:nm} and Table  \ref{tab:ks_mfof}).  The difference is less significant for the three--dimensional distance and this is likely caused by the relatively high peculiar velocities in the high mass groups that host LERGs.

The HERGs have closer nearest neighbours than their control samples, irrespective of the distance estimate used ($\simeq 99$ percent in both cases).  The most significant difference is found for the low luminosity HERGs, although there is no significant correlation between distance and radio luminosity.  As HERGs lie in lower mass groups, the influence of peculiar motions on the three--dimensional distance  $R_{\rm nm}$ should be less than for the LERGs.  These results suggest that both LERG and HERG activity may be enhanced interactions with other galaxies.

\section{Discussion}\label{sec:env_discussion}

It is increasingly clear that there are fundamental differences in the properties of radio galaxies with and without strong emission lines (i.e. HERGs and LERGs). The differences span a range of properties including stellar mass \citep[e.g.][]{kauffmann08,best12}, stellar population \citep[e.g.][]{herbert10} and environment \citep[e.g.][]{best04,tasse08,sabater13}, and cannot be simply explained by a unified model.

We have used data from the GAMA survey, with radio galaxies identified in the LARGESS sample, to explore the environments of HERGs and LERGs. The GAMA survey targets fainter galaxies than SDSS (Fig.\ \ref{fig:massVz}) and has high completeness \citep[98.5 percent][]{2015MNRAS.452.2087L}, even for galaxies in high density regions.  As a result we are able to probe lower halo masses and extend our environmental measures to higher redshifts.  Our radio galaxy sample is flux limited in the SDSS $i$-band and does not have any colour selection applied, unlike some previous surveys that used colour cuts to target specific sub-samples e.g. luminous red galaxies (LRGs) \citep{cannon06} or quasars \citep{croom04,croom09}.  This ensures that we sample a broad range of radio galaxies.

\subsection{The role of halo mass}

We have shown that the local density ($\Sigma_{5}$) and the fraction in groups are both significantly higher for high-luminosity LERGs than for their control sample. In contrast, the HERGs and low-luminosity LERGs show no significant difference in either the $\Sigma_{5}$ distribution or the fraction in groups when compared to their control samples. This confirms, at higher redshift (and with a larger sample), the results from \cite{sabater13} and \cite{best04}.

Our results imply that high-luminosity LERGs are preferentially found in more massive haloes than non-radio galaxies of similar stellar mass.  The high fraction of high-luminosity LERGs in groups also points towards a higher average halo mass for this population, because there is a minimum mass for the groups used in this analysis.  Galaxies hosted by lower mass halos would not be found in the group catalogue. Hence, both environmental measurements are proxies for the halo mass.

HERGs and low-luminosity LERGs, on the other hand, lie in halos that are similar in mass to non-radio galaxies of similar stellar mass. Because we make this comparison to stellar mass matched samples, the difference between HERGs and LERGs is not simply due to them having different stellar masses.

Cross-correlation studies of radio galaxies \cite[e.g.][]{wake08,mandelbaum09,donoso10} have shown that radio-loud LRGs (analogous to LERGs) are in more massive haloes than non-radio galaxies of similar stellar mass, which is in agreement with our results on $\Sigma_{5}$ and group fraction.  {\update Also in agreement with our work, \citet{2013MNRAS.430.3086G} use a sample of $z<0.3$ radio galaxies to show that LERGs have denser average environments than HERGs, although there is no direct control for stellar mass in their work.  \citet{2015MNRAS.453.2682I}  use a powerful alternative approach to show that the X-ray luminosity of groups and clusters hosting LERGs is positively correlated with radio luminosity, also supporting the environmental dependence of LERGs.}

The preference of high-luminosity LERGs for higher mass haloes is confirmed by a comparison to a matched sample that are also in GAMA groups.  Groups hosting high-luminosity LERGs are 0.2 dex higher in mass than the groups hosting control galaxies, and the difference is significant at the 98 percent level.  The difference is less (0.1 dex), but still significant (97 percent) for low-luminosity LERGs.  The difference in halo mass is a lower limit to the true difference, as in this measurement we only include those galaxies detected in GAMA groups and more control galaxies than LERGs lie in haloes below the GAMA group mass limits.

\subsection{The Triggering of LERGs}

High-mass haloes have cooling times that can be much longer than a dynamical time, so they are expected to have a smaller fraction of gas in a cold phase.  For example, in the simulations of \cite{keres09} haloes of mass of $10^{12} M_{\sun}$ have $\sim5$ percent of gas colder than 250,000 K, but this drops to $\sim0$ percent for a halo mass of $10^{13} M_{\sun}$. This is qualitatively consistent with observations: for example massive elliptical galaxies and FRI radio galaxies typically have relatively little H{\sc i} gas \citep{knapp78,morganti01,serra12}.  The lack of star formation in the LERG population also points to a deficit of cold gas and their stellar mass ($\sim10^{11} h^{-2} M_{\sun}$) is in the regime where cold gas accretion is not expected to be important \citep[e.g.][]{keres05}.

The supply of gas in LERGs could come from gas that is cooling from the virial temperature within the hot halo, and this is sufficient to power LERGs \citep{hardcastle07}. We find a marginal tendency (90 to 96 percent significant) for LERGs to be closer to the centre-of-light of a group, even when we control for them being the BCG or central galaxy.  Being in the centre of the group would also naturally lead to our discovery that LERGs are typically closer to their nearest neighbours than control galaxies.  If confirmed, this could imply that LERGs inhabit more dynamically evolved groups, where the radio-galaxy has had greater time to sink to the bottom of the gravitational potential via dynamical friction.  Such a location could lead to a higher accretion rate, or alternatively the higher density hot gas could provide a better working surface for the radio jet.  {\update \citet{2015MNRAS.453.2682I} argue that the correlation found between environment and luminosity in LERGs is more likely due to fuelling, as the luminosity boosting by providing a better working surface for the jet should be equally present in the HERG population, but is not seen.}

LERGs with higher radio luminosities are more strongly influenced by their environment, but environment still appears to play a role in lower luminosity objects.  For example, while there is no significant difference between low-luminosity LERGs and controls for $\Sigma_{5}$ or group fraction, there is a small but significant difference in group mass.  However, the fact that not all LERGs are in high density environments [the median $\log(\Sigma_{5})$ is $0.91\pm0.11$, but  12/75 LERGs have  $\log(\Sigma_{5})<0$] means that other factors must also play a role.  These could include galaxy-galaxy interactions \citep[e.g.][]{sabater13} driving small amounts of gas onto the nucleus, or internal stochastic gas accretion events.

{\update The increasing fraction of radio galaxies (dominated by LERGs) at high stellar mass [e.g.\ Fig.\ \ref{fig:pbest07}, and also see \citet{best07}] implies a higher duty cycle.  This makes it easier for LERGs to provide the maintenance--mode feedback in galaxy formation models and is consistent with the high fraction of radio galaxies in cool--core clusters \citep[e.g.][]{2006MNRAS.373..959D}.}

\subsection{The Triggering of HERGs}

HERGs are found in similar large-scale over-densities to non-radio galaxies matched in stellar mass and colour. When we compared HERGs and control galaxies that are both in groups, however, we found that the distance to the nearest group member is significantly shorter for HERGs than their control sample. This could be evidence that HERGs are influenced by galaxy-galaxy interaction. If so, it supports the claims \citep[e.g.][]{hardcastle07,best12} that HERGs might be triggered by mergers or interactions, which cause cold gas (from either the host galaxy or the interacting galaxy) to flow into the SMBH.  {\update \citet{2012MNRAS.419..687R} finds the fraction of luminous radio  galaxies (of which 76 percent are HERGs in their sample)  that show strong tidal features is significantly larger than for quiescent galaxies, supporting the merger/interaction triggering hypothesis.  The lower space density of HERGs \citep[e.g.][]{2016MNRAS.460....2P}, combined with their lower typical stellar mass (e.g.\ Fig.\ \ref{tab:f_group}), suggests that they have a much lower duty cycle than the LERG population.  This points to HERG activity being a transient phenomenon.}

One surprising result is that low luminosity HERGs have a much lower (by 1.4 dex) median group mass than their control sample.  These low luminosity HERGs also have nearest neighbours that are much closer (see Fig. \ref{fig:nm} and Table \ref{tab:ks_mfof})  In contrast, every other sample shows an increase in group mass compared to a control sample.  This could be because the low radio luminosity HERGs are contaminated by galaxies that are not true radio-AGN, but galaxies where star formation dominates the radio emission.  These contaminating galaxies may still show AGN signatures in the optical, but are dominated by star formation at radio wavelengths.  As discussed by \cite{2016MNRAS.460....2P}, there is no easy solution to this problem.  Approaches such as those taken by \cite{best12}, who use the strength of the D4000 spectral feature to discriminate between star formation and AGN, lead to the rejection of large numbers of HERGs which have bluer (and therefore younger) stellar populations than LERGs \citep{herbert10}. {\update We have inspected the radio morphology of the low luminosity HERGs to see if they show evidence of the radio emission being due to a star--forming disk, however all but one of these objects is a point source in FIRST.  One galaxy does have extended flux in the radio, but with an orientation that is offset by $\sim45^\circ$ from the optical major axis of the galaxy.  Therefore, from the radio morphology it is inconclusive whether the HERGs are contaminated by star--forming galaxies.}

\subsection{Dependence on galaxy morphology}

If the radio galaxies and controls have a different morphological mix the above results could be due to morphology, rather than radio emission.  \cite{worpel13} studied the environment (via the two-point cross-correlation and counts-in-cylinders) of a sample of local elliptical radio galaxies (selected using colour and visual morphology) from the \cite{mauch07} sample.  They compare the radio galaxies to a control sample matched in redshift, colour, magnitude and morphology drawn from the 6-degree Field Galaxy Survey \citep[6dFGS;][]{jones04}.  Worpel et al.\ find no statistical difference between radio galaxies and controls using the two-point cross-correlation function, in apparent conflict with other studies \citep[e.g.][]{wake08,donoso10}.   \cite{worpel13} argue that this may be driven by contamination from late-type galaxies in other samples.  As a result, we should consider whether host morphology can influence our results. 

We repeat our \emph{fraction-in-group} analysis with disk-like galaxies removed from both the LERGs and controls.  As a morphology proxy we use the concentration index $C=R_{90}/R_{50}$, where $R_{90}$ and $R_{50}$ are the radii containing 90\% and 50\% of the $r$-band Petrosian flux respectively\citep{strateva01,shimasaku01}.  Values of $C\sim3.0$ and  $C\sim2.3$ correspond to de Vaucouleurs (early-type) and exponential disk (late-type) profiles respectively.  We repeat our measurements using only galaxies with $C > [2.3, 2.5, 2.7]$ and find that in each case there is a significantly higher fraction ($>2\sigma$) of LERGs in groups than the controls.  For $C>3.0$ the significance is $<2\sigma$ as the sample size is reduced to $\simeq1/3$ (or $\sim100$ galaxies).

As a second test for the impact of morphology we add $|\Delta C| <0.1$ as an additional parameter for selecting control galaxies.  Again, we still find that a higher fraction of LERGs are in groups than their controls, although the significance drops from $3.5\sigma$ to $2.7\sigma$.

We conclude that morphology is not the main factor that produces the observed differences between radio galaxies and their controls.  We note that when Worpel et al.\ takes a deeper look at the environment around their radio galaxies (by using SDSS imaging) they find tentative evidence for larger numbers of satellite galaxies around their radio-galaxies than their controls.  This suggests that observations which are deeper than the typical magnitude of the radio galaxy (such as that available in our GAMA sample) are important for clearly discriminating environmental trends.

\section{Summary}\label{sec:env_summary}

Using a sample of LARGESS radio galaxies in the GAMA survey, we have conducted a detailed study of environment for high- and low- excitation radio galaxies at redshifts up to 0.4. In addition to the redshift range, other advantages of our sample compared to previous studies is the base GAMA sample and the radio galaxy selection method. The base GAMA galaxy sample provides a large, deep and complete sample of galaxies to measure galaxy environments and provide control samples. As with the radio galaxy sample from \cite{best12}, our radio galaxies are selected without any colour selection (unlike other samples e.g. \citealt{sadler07,donoso09}) and enables us to have a more complete range of radio galaxies.

\subsection{The environments of low-excitation radio galaxies (LERGS)}

Our results suggest that the high-luminosity radio galaxies with weak or no emission lines (LERGs) lie in more massive haloes than non-radio galaxies of similar stellar mass and colour, which is consistent with previous studies \citep[e.g.][]{best04,wake08,mandelbaum09,donoso10}. We do not see this difference in halo mass for low-luminosity LERGs, except when we only compare those in groups, in which case, low-luminosity LERGs are also in higher mass haloes than their control sample.

Once we control for stellar mass, colour and group membership, LERGs lie slightly closer to the centre of their group than non-radio galaxies. There is weak evidence that this difference in radial distribution is due to the most central galaxy (which we have defined as the iterative central galaxy or IterCen) and possibly BCGs. LERGs that are the IterCen and BCGs of their groups, lie slightly closer to the group centre than non-radio IterCen galaxies and BCGs, but this difference is only marginally significant. Together these facts support the idea \cite[e.g.][]{best04} that the environment plays a role in triggering or providing a better working surface for the jets of LERGs.

\subsection{The environments of high-excitation radio galaxies (HERGs)}

Radio galaxies with strong optical emission lines (HERGs) have similar environments to non-radio galaxies of equal stellar mass and colour.  The HERGs are typically in lower mass haloes than LERGs, consistent with them also having lower stellar masses than LERGs. Previous studies \citep[e.g.][]{best04,sabater13}, using lower redshift samples, have also shown that LERGs have a preference for high mass haloes, while such a dependence is absent in HERGs. Within galaxy groups, we find that HERGs have a closer nearest neighbours and lower halo masses than their controls, but that this is dominated by HERGs with low radio luminosity.  We argue that this result could be influenced by contamination by galaxies where the radio emission is dominated by star formation.

\section*{Acknowledgements}

JHYC would like to acknowledge the funding provided by the University of Sydney via the Australian Postgraduate Award, the Postgraduate Research Scholarship Scheme and the William \& Catherine McIlrath Scholarship. JHYC would also like to acknowledge the support provided by the Astronomical Society of Australia. EMS and SMC acknowledge the financial support of the Australian Research Council under the grants DP1093086 and DP130103198. SMC acknowledges the support of an Australian Research Council Future Fellowship (FT100100457). We thank Philip Best for valuable input into this work.

GAMA is a joint European-Australasian project based around a spectroscopic campaign using the Anglo-Australian Telescope. The GAMA input catalogue is based on data taken from the Sloan Digital Sky Survey and the UKIRT Infrared Deep Sky Survey. Complementary imaging of the GAMA regions is being obtained by a number of independent survey programs including GALEX MIS, VST KIDS, VISTA VIKING, WISE, Herschel-ATLAS, GMRT and ASKAP providing UV to radio coverage. GAMA is funded by the STFC (UK), the ARC (Australia), the AAO, and the participating institutions. The GAMA website is http://www.gama-survey.org/.

Funding for the SDSS and SDSS-II has been provided by the Alfred P. Sloan Foundation, the Participating Institutions, the National Science Foundation, the U.S. Department of Energy, the National Aeronautics and Space Administration, the Japanese Monbukagakusho, the Max Planck Society, and the Higher Education Funding Council for England. The SDSS Web Site is http://www.sdss.org/.

The SDSS is managed by the Astrophysical Research Consortium for the Participating Institutions. The Participating Institutions are the American Museum of Natural History, Astrophysical Institute Potsdam, University of Basel, University of Cambridge, Case Western Reserve University, University of Chicago, Drexel University, Fermilab, the Institute for Advanced Study, the Japan Participation Group, Johns Hopkins University, the Joint Institute for Nuclear Astrophysics, the Kavli Institute for Particle Astrophysics and Cosmology, the Korean Scientist Group, the Chinese Academy of Sciences (LAMOST), Los Alamos National Laboratory, the Max-Planck-Institute for Astronomy (MPIA), the Max- Planck-Institute for Astrophysics (MPA), New Mexico State University, Ohio State University, University of Pittsburgh, University of Portsmouth, Princeton University, the United States Naval Observatory, and the University of Washington.

\bibliographystyle{mn2e}

\bibliography{phd_lit.bib,journal_abb}


\label{lastpage}
\end{document}

%% file: authors.tex
\author[J.~H.~Y.~Ching et al.]
{{\parbox{\textwidth}{\raggedright J.~H.~Y.~Ching$^{1}$,
S.~M.~Croom$^{1}\thanks{E-mail:~scott.croom@sydney.edu.au}$,
E.~M.~Sadler$^{1}$,
A.~S.~G.~Robotham$^{2,3}$,
S.~Brough$^{4}$,
I.~K.~Baldry$^{5}$,
J.~Bland-Hawthorn$^{1}$,
M.~Colless$^{6}$,
S.~P.~Driver$^{2,3}$,
B.~W.~Holwerda$^{7}$
A.~M.~Hopkins$^{4}$,
M.~J.~Jarvis$^{8,9}$,
H.~M.~Johnston$^{1}$,
L.~S.~Kelvin$^{2,3,5,10}$,
J.~Liske$^{11}$,
J.~Loveday$^{12}$,
P.~Norberg$^{13}$,
M.~B.~Pracy$^{1}$,
O.~Steele$^{14}$,
D.~Thomas$^{14}$
L.~Wang$^{15,16}$
}}\\
\vspace{0.4cm}\\
{\parbox{\textwidth}{\raggedright 
$^{1}$Sydney Institute for Astronomy, School of Physics, University of Sydney, NSW 2006, Australia
\\
$^{2}$Scottish Universities Physics Alliance (SUPA), School of Physics, University of St Andrews, St Andrews, Fife, UK, KY169SS
\\
$^{3}$International Centre for Radio Astronomy Research (ICRAR), University of Western Australia, Crawley, WA6009, Australia
\\
$^{4}$Australian Astronomical Observatory, P.O.~Box~915, North Ryde, NSW 1670, Australia
\\
$^{5}$Astrophysics Research Institute, Liverpool John Moores
University, Twelve Quays House, Egerton Wharf, Birkenhead, CH41 1LD,
UK
\\
$^{6}$Research School of Astronomy and Astrophysics, The Australian National University, Canberra 2611, Australia
\\
$^{7}$Department of Physics and Astronomy, University of Louisville,
Louisville KY 40292 USA 
\\
$^{8}$Astrophysics, Department of Physics, University of Oxford, Keble Road, Oxford, OX1 3RH
\\
$^{9}$Physics Department, University of the Western Cape, Private Bag X17, Bellville 7535, South Africa
\\
$^{10}$Institut f\"{u}r Astro- und Teilchenphysik, Universit\"{a}t Innsbruck, Technikerstra{\ss}e 25, 6020 Innsbruck, Austria
\\
$^{11}$Hamburger Sternwarte, Universit\"{a}t Hamburg, Gojenbergsweg
112, 21029 Hamburg, Germany
\\
$^{12}$Astronomy Centre, University of Sussex, Falmer, Brighton BN1 9QH, UK
\\
$^{13}$Institute for Computational Cosmology, Department of Physics, Durham University, South Road, Durham DH1 3LE, UK
\\
$^{14}$Institute of Cosmology and Gravitation (ICG), University of Portsmouth, Dennis Sciama Building, Burnaby Road, Portsmouth PO1 3FX, UK
\\
$^{15}$SRON Netherlands Institute for Space Research, Landleven 12,
 9747 AD, Groningen, The Netherlands
\\
$^{16}$Kapteyn Astronomical Institute, University of Groningen,
Postbus 800, 9700 AV Groningen, the Netherlands
\\
}}}